\documentclass[useAMS]{mn2e}

\usepackage[dvips]{graphicx}
\usepackage{amssymb}
\usepackage[english]{babel}

\newcommand{\Msol}{$M_\odot$}

\newcommand{\lsim}{\raisebox{-3.8pt}{$\;\stackrel{\textstyle <}{\sim}\;$}}
\newcommand{\DYDZ}{\Delta Y/\Delta Z}
\newcommand{\feh}{\rm{[Fe/H]}}
\newcommand{\mh}{\rm{[M/H]}}
\newcommand{\rMS}{\rm{rMS}}
\newcommand{\bMS}{\rm{bMS}}
\newcommand{\mMS}{\rm{mMS}}
\newcommand{\teff}{T_{\rm{eff}}}
\newcommand{\etal}{\mbox{{\rm et~al.\ }}}

\title[$\DYDZ$ in Globular Clusters: insight from nearby stars]
{Revisiting $\DYDZ$ from multiple main sequences in Globular Clusters: insight 
from nearby stars}

\author[Portinari, Casagrande and Flynn]{Laura Portinari$^{1}$, 
Luca Casagrande$^{2}$ and Chris Flynn$^{1,3}$\thanks{E-mail: 
{\tt lporti,cflynn@utu.fi, luca@mpa-garching.mpg.de}}\\
$^1$ Tuorla Observatory, Department of Physics and Astronomy, University 
of Turku, V\"ais\"al\"antie 20, FIN-21500 Piikki\"o, Finland\\
$^2$ Max Planck Institute for Astrophysics, Karl Schwartzschild Stra\ss e 1,
Garching, Germany\\
$^3$ FINCA -- Finnish Centre for Astronomy with ESO, University 
of Turku, V\"ais\"al\"antie 20, FIN-21500 Piikki\"o, Finland}
\begin{document}
\date{Accepted 2010 April 7. Received 2010 April 6; in original form 2009 December 7}

\maketitle

\begin{abstract}
For nearby K dwarfs, the broadening of the observed Main Sequence at low 
metallicities is much narrower than expected from isochrones with the standard 
helium--to--metal enrichment ratio $\DYDZ \sim 2$. Though the latter value fits 
well the Main Sequence around solar metallicity, and agrees with independent 
measurements from HII regions as well as with theoretical stellar yields
and chemical evolution models, a much higher $\DYDZ\sim10$ is necessary 
to reproduce the broadening observed for nearby subdwarfs. 
This result resembles, on a milder scale, 
the very high $\DYDZ$ estimated from the multiple Main Sequences 
in $\omega$Cen and NGC~2808. 
Although not ``inverted'' as in $\omega$Cen, where the metal-rich Main 
Sequence is bluer than the metal-poor one, the broadening observed for nearby 
subdwarfs is much narrower than stellar models predict for a standard 
helium content.
We use this empirical evidence to argue that a revision of lower Main Sequence 
stellar models, suggested from nearby stars, could significantly reduce 
the helium content inferred for the subpopulations of those globular clusters. 
A simple formula based on empirically calibrated homology relations is 
constructed, for an alternative estimate of $\DYDZ$ in multiple main 
sequences. We find that, under the most favourable assumptions, the estimated
helium content for the enriched populations could decrease from $Y \simeq 0.4$ 
to as low as $Y \simeq 0.3$.
\end{abstract}

\begin{keywords}
stars: abundances - stars: fundamental parameters -
stars: subdwarfs - Hertzsprung--Russell (HR) diagram - globular
clusters: individual: $\omega$ Cen, NGC 2808
\end{keywords}

%%%%%%%%%%%%%%%%%%%%%%%%%%%%%%%%%%%%%%%%%%%%%%%%%%%%%%%%%%%%%%%%%%%%%%%%%%%%%%%
\section{Introduction}\label{sec:intro}

Helium is the second most abundant element in the Universe, having been
produced by Big Bang Nucleosynthesis (BBN) with a universal mass fraction of 
$Y_P \sim$ 0.24 (Steigman 2007); after that, successive stellar generations 
synthesize metals ($Z$) and an additional amount of helium $\Delta Y$. The 
characteristic helium--to--metal enrichment rate $\DYDZ$ is estimated to be 
around 2, both observationally from HII regions (e.g.\ Fukugita 
\& Kawasaki 2006; Peimbert \etal 2007; Izotov, Thuan \& Stasinska 2007) 
and theoretically, from models of 
stellar nucleosynthesis and chemical evolution of the Galactic disc
(e.g.\ Chiosi \& Matteucci 1982; Maeder 1992; Carigi \& Peimbert 2008; 
Casagrande 2008)

In spite of its large abundance, helium is an elusive element. 
It can be measured directly, from spectroscopic lines, only in stars
hotter than $\sim 10,000$~K: young, massive stars (or their surrounding 
HII regions)
or blue Horizontal Branch stars (where though, apart from a small temperature 
range, the surface abundance may not trace the original one, due to helium 
sedimentation and metal levitation: Michaud, Vauclair \& Vauclair 1983; 
Michaud, Richer \& Richard 2008; Villanova, Piotto \& Gratton 2009 
and references therein).
Only indirect methods can be used for stars of lower
mass and cooler temperatures, which constitute the bulk of the Galaxy's stellar
population. One such method relies on the broadening
of the lower Main Sequence (MS): in fact the location of low--mass, long--lived 
stars 
in the HR diagram depends on their metallicity $Z$ and their helium 
content $Y$  --- or equivalently, on $\DYDZ$. Concomitant increases in $Z$ and 
$Y$ shift the MS in opposite directions, so for a given variation $\Delta Z$
the two MSs are more spread apart (the metal rich being cooler at a given 
luminosity) if the corresponding $\Delta Y$ is lower; conversely, 
if $\Delta Y$ is large ``enough'', the MSs may even invert, with the more metal
(and helium) rich being bluer (i.e.\ hotter; e.g.\ Fernandes \etal 1996).

This method dates back to Faulkner (1967); early studies based on ground--based
parallaxes deduced a large and quite uncertain value of $\DYDZ\sim 5 \pm 3$
from the apparent overlap of MSs of all metallicities (e.g.\ Perrin \etal 
1977; Fernandes \etal 1996). The situation greatly improved with Hipparcos 
parallaxes, so that a net separation of the MS as a function
of $Z$ could be detected in the HR diagram. Studies based on Hipparcos 
distances concluded $\DYDZ=2-3$ (Pagel \& Portinari 1998; Jimenez \etal 2003).
More recently, Casagrande \etal (2006, 2007) further improved the analysis 
by compiling a much larger sample of K dwarfs with homogeneous multi--band
photometry, for which bolometric magnitudes and effective temperatures 
were derived with the InfraRed Flux Method (IRFM).
The interpretation of the 
broadening of the lower MS could thus be performed in the theoretical HR diagram
($M_{\rm{Bol}}$ vs.\ $\log \teff$) where comparison to stellar models is more
straightforward and the effects of $\DYDZ$ more prominent (Castellani,
Degl'Innocenti \& Marconi 1999). Around solar metallicities this analysis 
yielded $\DYDZ \sim2$; but at lower $Z$ Casagrande \etal (2007) 
found that the observed Main Sequence is narrower than expected from 
stellar models computed under the standard assumption $\DYDZ \sim 2$, 
thus implying a very steep $\DYDZ \sim 10$ (Fig.~\ref{fig:YZfield}). 
This change in slope is at odds
with Galactic chemical evolution models, that predict a ratio substantially 
constant with $Z$ (Carigi \& Peimbert 2008; Casagrande 2008); a constant
$\DYDZ \sim 2$ is also found for both metal-rich (e.g.\ Peimbert 2003; Balser 
2006) and metal-poor (e.g.\ Fukugita \& Kawasaki 2006; Peimbert \etal 2007;
Izotoz \etal 2007) HII regions. 
More importantly, $\DYDZ \sim 10$ at low metallicities implies a helium
content for nearby subdwarfs much lower than $Y_P$, in awkward constrast 
with the cosmological floor set by standard BBN. 

A combination of small sample size and $Z$--dependent bolometric corrections
in the observational HR diagram had masked this effect in previous studies 
of $\DYDZ$, although very low $Y$ values had already been noticed in a handful
of low metallicity stars with available IRFM $M_{\rm{bol}}$ and $\teff$ 
(Fernandes \etal 1998; Lebreton \etal 1999).
 This result points toward inadequacies in Main Sequence stellar models
of low metallicity, and we now plan to investigate the issue further, 
as this may lead to reconsideration of the problem of the helium enrichment 
in globular clusters.  

\begin{figure}
\begin{center}
\includegraphics[scale=0.43]{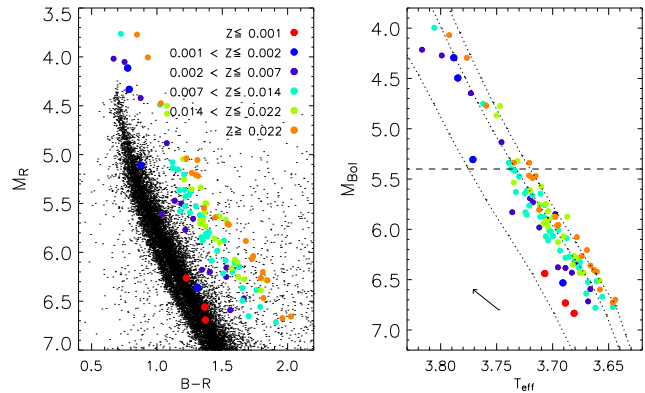}
\caption{{\it Left panel}: HR diagram of nearby K dwarfs (filled circles) 
from the sample of Casagrande \etal (2007), overplotted on the MSs photometry 
of $\omega$Cen (Sollima \etal 2007), adopting $E(B-V)=0.11$ (Lub 2002) and 
dereddened distance modulus of 13.7 (Bellazzini \etal 2004; Del Principe 
\etal 2006). The maximum split between the red and blue MS of $\omega$Cen 
$\Delta (B-R)=0.1$ occurs around $M_R \sim M_{\rm{Bol}} \sim 6.5$ (Sollima 
\etal 2007). {\it Right panel}: nearby K dwarfs in the theoretical plane 
with isochrones having $Z=0.001, Z_{\odot}, 0.04$ and $\Delta Y / \Delta Z=2$
overplotted. Clear mismatch in the broadening appears for low metallicities. 
Below $M_{\rm{Bol}}=5.4$ the broadening of the MS is age independent.
We also show the reddening/extinction correction corresponding
to $E(B-V)=0.10$ that would be needed to reconcile the metal-poor stars 
with theoretical isochrones.}
\label{fig:WCen}
\end{center}
\end{figure}

In fact, at absolute magnitudes comparable to those of the local stars 
studied in Casagrande \etal (2007, $M_V \sim 5.5-7.5$), multiple Main Sequences 
have been discovered in some globular clusters and interpreted as evidence 
for huge helium enhancement in a sub-population (Bedin \etal 2004; Norris 
2004; Piotto \etal 2005, 2007); see Fig.~\ref{fig:WCen}. The required 
$\DYDZ \gtrsim 100$ is extremely difficult to explain with stellar 
nucleosynthesis and chemical evolution models (Karakas \etal 2006; Prantzos 
\& Charbonnel 2006; Maeder \& Meynet 2006; Choi \& Yi 2007, 2008; Romano 
\etal 2007, 2009; Renzini 2008; Yi 2009; Marcolini \etal 2009; 
Peng \& Nagai 2009). 

The helium content of the sub-population also has a major role in shaping
other regions of the HR diagram (Catelan, Valcarce \& Sweigart 2009a 
and references therein); in particular it is reflected
in the morphology of the Horizontal Branch  (D'Antona \& Caloi 2004; 
Caloi \& D'Antona 2005, 2007; Lee \etal 2005).
While it is likely that a significant helium enhancement is present in those 
stellar populations, the purpose of this work is to show that the  
revision of low metallicity MS stellar models, needed to cure the problem 
of the high $\DYDZ$ and sub-primordial $Y$ deduced in nearby K dwarfs, may 
significantly reduce current estimates of $\DYDZ$ in globular clusters with 
multiple MSs, easing their theoretical interpretation.

We show this by means of an exercise based on homology relations.
Theoretical homology relations, describing how the location of the low MS
in the HR diagram depends on metallicity and helium content,
are briefly recalled in Section~\ref{sec:homology}. In Section~3, 
we apply homology relations to interpret the split of multiple MSs
in globular clusters, and show that the inferred helium enrichment
for the blue population(s) agrees with the results of detailed isochrone
analysis. In Section~4 we confirm that, in general, homology relations
render very well the behaviour of stellar models as a function of the 
helium content. However, both isochrones and homology relations fail
the interpretation of the low MS of nearby stars ---  as we cannot accept
sub--primordial helium contents for low metallicity stars. Therefore,
in Section~5 we proceed to calibrate {\it empirical} homology relations,
that return the expected $\DYDZ = 2$ for the low MS of nearby stars
with $Z<Z_{\odot}$.
Such empirically calibrated homology relations also return, for the
helium--rich subpopulations of globular clusters, a helium content 
$Y \sim 0.3$, rather than the much higher $Y \sim 0.35-0.4$ of standard
analysis. This lower value of helium enrichment can be reconciled
far more easily with present chemical evolution models (Karakas \etal 2006;
Renzini 2008; Yi 2009).
Finally, we conclude in Section~6 with a discussion of the possible physical
processes that could improve stellar models for low MS stars, so as to
solve the ``helium problem'' both for nearby stars and for globular clusters.

%%%%%%%%%%%%%%%%%%%%%%%%%%%%%%%%%%%%%%%%%%%%%%%%%%%%%%%%%%%%%%%%%%%%%%%%%%%%%%
%%%%%%%%%%%%%%%%%%%%%%%%%%%%%%%%%%%%%%%%%%%%%%%%%%%%%%%%%%%%%%%%%%%%%%%%%%%%%%%

%%%%%%%%%%%%%%%%%%%%%%%%%%%%%%%%%%%%%%%%%%%%%%%%%%%%%%%%%%%%%%%%%%%%
\begin{figure*}
\begin{center}
\includegraphics[scale=0.43]{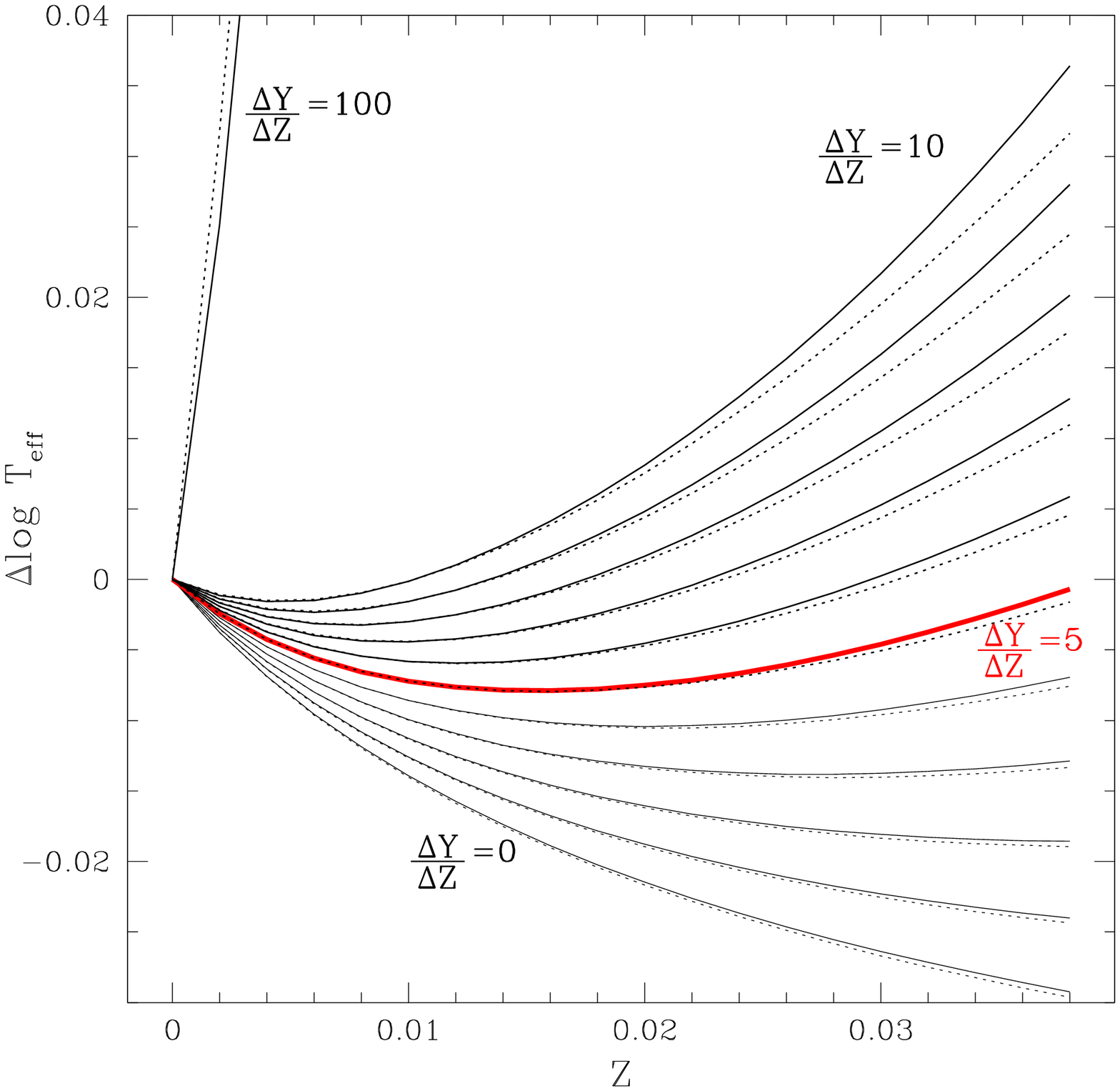}
\includegraphics[scale=0.43]{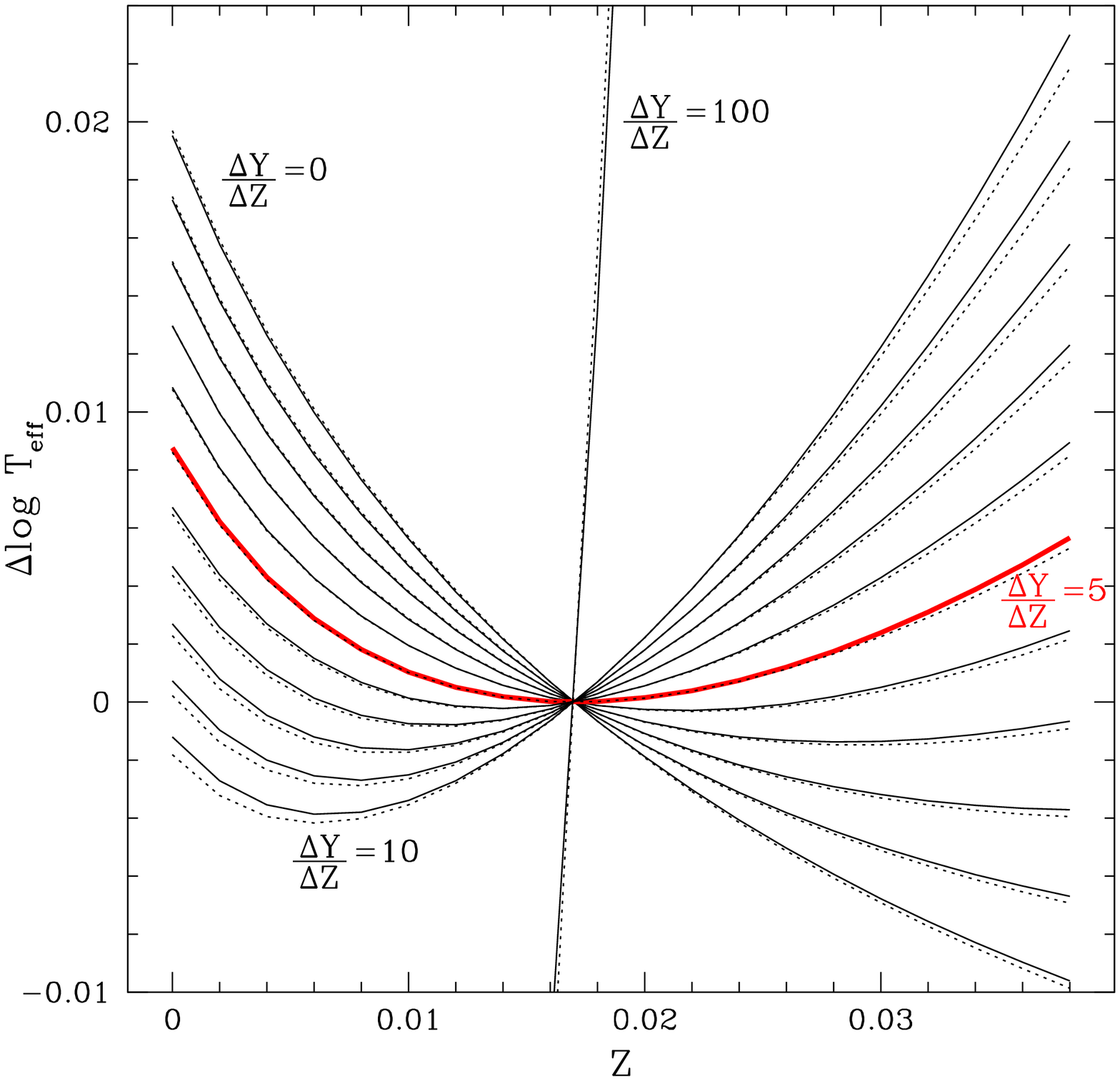}
\caption{Broadening of the ZAMS in log $\teff$ as a function of metallicity,
as predicted by homology relations (Eq.~\protect{\ref{eq:DlogTeff}}).
The broadening is referred to the 
primordial ZAMS ($Z_P$, $Y_P$; left panel) or to the solar ZAMS 
($Z_\odot=0.017$, $Y_\odot=0.263$; right panel) for $\frac{\Delta Y}{\Delta Z}$ 
ranging from 0 to 10 as indicated. Dotted lines show the approximate relation 
proposed in Eq.~\protect{\ref{eq:DlogTeff_2terms}}.
The thick red line highlights the case $\frac{\Delta Y}{\Delta Z}=5$, which 
corresponds to MS inversion around $Z_\odot$. At increasing $\DYDZ$, MS 
inversion occurs at lower and lower $Z$.}
\label{fig:homology}
\end{center}
\end{figure*}
%%%%%%%%%%%%%%%%%%%%%%%%%%%%%%%%%%%%%%%%%%%%%%%%%%%%%%%%%%%%%%%%%%%%%%%%%

%%%%%%%%%%%%%%%%%%%%%%%%%%%%%%%%%%%%%%%%%%%%%%%%%%%%%%%%%%%%%%%%%%%%%%%%%%%%%
\section{Homology relations}\label{sec:homology}

Homology relations (Cox \& Giuli 1968), holding for radiative structures
like the energy-producing cores of MS stars of $M$\lsim 1~\Msol, express
the difference in $M_{\rm{Bol}}$ of a Zero Age Main Sequence of composition
($Y,Z$) with respect to another reference ZAMS of composition 
($Y_r,Z_r$). In the hypothesis that the compositions are related by:
\[ Y= Y_r + \frac{\Delta Y}{\Delta Z} (Z-Z_r) \]
the formula is:
\begin{equation}
\label{eq:DMbol} 
\begin{array}{l l}
\Delta M_{\rm{Bol}} = 

\medskip
     & -1.59 \log \left[ 1-\frac{\delta}{X_r} (Z-Z_r) \right] \\

\medskip
    & - 3.33 \log \left[ 1-\frac{5 \delta +1}{(3+5 X_r-Z_r)}(Z-Z_r) \right] \\

\medskip
            &  -0.8675 \log \left[ 1-\frac{\delta}{(1+X_r)} (Z-Z_r) \right] \\
      &  -0.8675 \log \left( \frac{100 Z +1}{100 Z_r +1} \right) \\
\end{array}
\end{equation}
where $\delta=1+ \frac{\Delta Y}{\Delta Z}$ and $X_r=1-Y_r-Z_r$. Notice that 
the same relation holds also for a generic combination of $(Y,Z)$ replacing 
$\delta \, (Z-Z_r) \rightarrow (Y-Y_r)+(Z-Z_r)$.
The first term is related to the rate of thermonuclear energy generation,
the second term to the molecular weight, and the final two terms to opacity
(Cox \& Giuli 1968; Fernandes et~al.\ 1996; see also the Appendix). For 
$Z_r=0$,  $Y_r=Y_P$ (primordial helium floor due to BBN) this formula 
corresponds to eq.~3 of Pagel \& Portinari
(1998). Another typical approach is to refer to the solar 
ZAMS, as the solar model is the basic calibration of stellar tracks and 
isochrones; then ($Y_r,Z_r$)=($Y_\odot,Z_\odot$).

In the case of old stellar populations, because evolution affects luminosities 
approximately brighter than $M_{\rm{Bol}}$=5.4, it is more useful to translate 
the broadening in $M_{\rm{Bol}}$ into a broadening in $\teff$ using the slope 
of the lower MS.
In the Padova isochrones used in Casagrande \etal (2007), this
slope is about 17 mag per dex in log($\teff$), henceforth:
\begin{equation}
\label{eq:DlogTeff}
\begin{array}{l l}
\Delta \log \teff= 

\medskip
                 & -0.0935 \log \left[ 1-\frac{\delta}{X_r} (Z-Z_r) \right]  \\

\medskip
      & -0.196 \log \left[ 1-\frac{5 \delta +1}{(3+5 X_r-Z_r)}(Z-Z_r) \right] \\

\medskip
           &  -0.051 \log \left[ 1-\frac{\delta}{(1+X_r)} (Z-Z_r) \right] \\
                  & -0.051 \log \left( \frac{100 Z+1}{100 Z_r+1}  \right) \\

\end{array}
\end{equation}                
(cf.\ eq.~4 of Pagel \& Portinari, with slightly different coefficients 
due to the different adopted slope of the MS). 
Fig.~\ref{fig:homology} shows the broadening of the MS as a function of 
metallicity,
as predicted by homology relations. For low $\DYDZ$ (e.g.\ $\DYDZ=0$) the MS
is cooler at increasing metallicity, as ``normally'' expected. When $\DYDZ=5$,
MS inversion is expected around solar metallicity (i.e.\ overlapping MS, 
as was deduced from pre--Hipparcos data). At increasing $\DYDZ$, MS inversion 
--- namely, more metal-rich MS being hotter due to the overwhelming
effect of the helium excess $\Delta Y$ --- occurs at lower and lower $Z$.
Indeed, extremely high $\DYDZ$ are necessary to interpret the MS inversion
at the low $Z$ of $\omega$ Centauri.

%%%%%%%%%%%%%%%%%%%%%%%%%%%%%%%%%%%%%%%%%%%%%%%%%%%%%%%%%%%%%%%%%%%%

\section{Homology relations and globular clusters}
\label{sec:homology_GC}

Evidence for multiple stellar populations in some Globular Clusters comes 
from turn-off, subgiant and red giant branch splits; bi- or multi-modal 
distribution of light elements; and the presence of multiple, distinct MSs 
(e.g.\ Piotto 2009). 
In the latter case, because of the faintness of the stars, deep photometric 
observations are needed and clear evidence for multiple MSs is currently 
limited to $\omega$Cen and NGC~2808; in addition, evidence for MS 
broadening has been recently found in 47 Tuc (Anderson \etal 2009).

In this section we analyze the multiple MSs of the above mentioned clusters
by means of homology relations, and compare the results to those obtained
with modern isochrone fitting.

\subsection{$\omega$ Cen}
Detailed isochrone analysis indicates that a helium enrichment of 
$\Delta Y \simeq 0.15$ between the two populations is needed to reproduce
the inverted MSs of $\omega$Cen (Norris 2004; Piotto \etal 2005; Lee \etal 2005;
Sollima \etal 2007). We show here that homology relations yield a very similar
result. 
To apply homology relations (Eq.~\ref{eq:DlogTeff}), we must translate 
into a split in log $\teff$
the maximum observed colour split between the blue (bMS) and red (rMS) 
main sequence: $\Delta (B-R)=0.1$, which 
occurs at a dereddened $B-R=1.2$ for the rMS (Sollima \etal 2007).
We apply the colour--temperature--metallicity scale of Casagrande \etal 
(2006) that was derived for nearby K dwarfs, adopting $\feh_{\rMS}=-1.6$ and 
$\feh_{\bMS}=-1.3$ (Sollima \etal 2007).
We obtain for the rMS $\teff$=5120~K and $\Delta \log \teff = 0.0185$ 
between the two sequences. Considering possible uncertainties in the 
reddening which affect the absolute $B-R$ values (while the 
differential $\Delta (B-R)=0.1$ is robust) we estimate 
$\Delta \log(\teff) = 0.0185 \pm 0.0015$. Adopting the updated and
extended temperature scale by Casagrande \etal (2010) also yields 
$\Delta \log \teff$ estimates within this range.

%%%%%%%%%%%%%%%%%%%%%%%%%%%%%%%%%%%%%%%%%%%%%%%%%%%%%%%%%%%%%%%%%%%%
\begin{figure}
\includegraphics[scale=0.43]{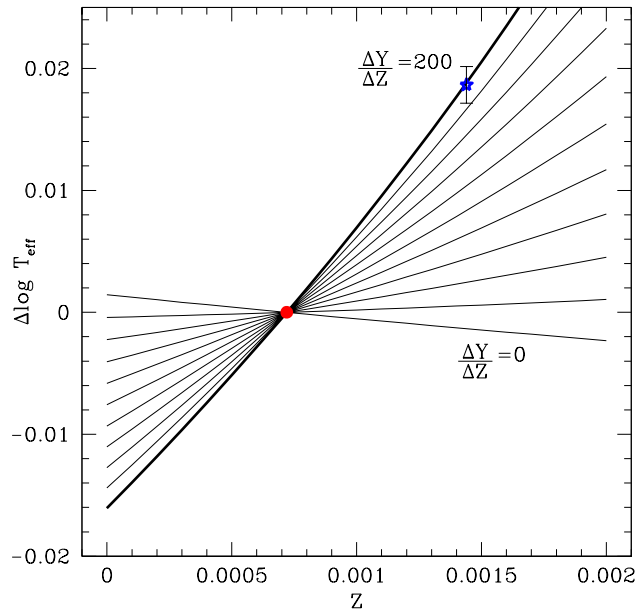}
\caption{Temperature split of the MS predicted by homology relations 
compared to the split between the rMS (dot, taken as reference MS 
in Eq.~\protect{\ref{eq:DlogTeff}}) and the bMS (star symbol) 
of $\omega$Cen. The observed temperature split is $\Delta \log(\teff) = 
0.0185 \pm 0.0015$, and the assumed metallicities are $Z_{\rMS}=0.00072$ and 
$Z_{\bMS}=0.00144$ (see text). Lines indicate homology predictions for 
$\DYDZ$ increasing from 0 to 200 in steps of 20.
The observed split between the rMS and bMS is reproduced for 
$\DYDZ \simeq 200$, corresponding to  $\Delta Y = 0.144$ and 
$Y_{\bMS} = 0.39$.}
\label{fig:WCen_homology}
%\end{center}
\end{figure}
%%%%%%%%%%%%%%%%%%%%%%%%%%%%%%%%%%%%%%%%%%%%%%%%%%%%%%%%%%%%%%%%%%%%

%%%%%%%%%%%%%%%%%%%%%%%%%%%%%%%%%%%%%%%%%%%%%%%%%%%%%%%%%%%%%%%%%%%%
\begin{figure}
\includegraphics[scale=0.43]{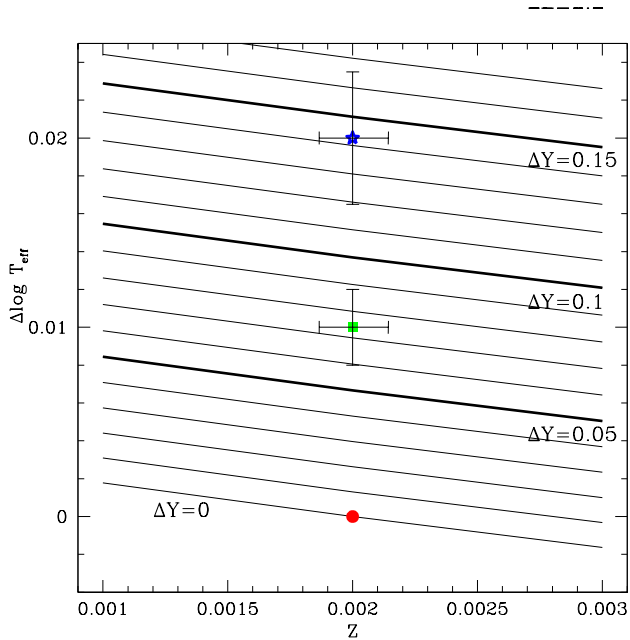}
\caption{Temperature split of the MS predicted by homology relations 
compared to the split between the red MS (dot, taken as reference MS 
in Eq.~\protect{\ref{eq:DlogTeff}}), the middle MS (square symbol) 
and the blue MS (star symbol) of NGC~2808. In this figure homology
relations are marked as a function of $\Delta Y$ rather than $\DYDZ$
(see text), with $\Delta Y$ increasing from 0 to 0.17 in steps of 0.01.
The observed splits are reproduced for $\Delta Y$(mMS-rMS)=0.074 and
$\Delta Y$(bMS-rMS)=0.143, corresponding to $Y_{\mMS} = 0.322$ and 
$Y_{\bMS} = 0.391$. Errors bars are estimated converting in 
$\Delta \log \teff$ an uncertainty of 0.02 magnitudes for the value of 
$\Delta(B-I)$.}
\label{fig:ngc2808_homology}
\end{figure}
%%%%%%%%%%%%%%%%%%%%%%%%%%%%%%%%%%%%%%%%%%%%%%%%%%%%%%%%%%%%%%%%%%%%

To derive the metal mass fraction $Z$ that is relevant for homology relations,
we adopt for both sequences an enrichment $\rm{[\alpha/Fe]}=+0.3$ as supported 
by various spectroscopic studies (Sollima \etal 2007 and references therein; 
Villanova \etal 2007). Using the relation of Yi \etal (2001) to compute the 
global metallicity leads to $\mh_{\rMS}=-1.37$ and $\mh_{\bMS}=-1.07$. 
Simple scaling with $Z_\odot$ returns $Z_{\rMS}=0.00072$ and $Z_{\bMS}=0.00144$. 
We further assume $Y_{\rMS}= 0.246$ --- but note that the resulting 
$\Delta Y= Y_{\bMS}-Y_{\rMS}$ is not sensitive to the adopted value of $Y_{\rMS}$.
Fig.~\ref{fig:WCen_homology} shows the split in $\Delta \log \teff$
predicted by homology relations with respect to the rMS, i.e.\
applying Eq.~\ref{eq:DlogTeff} with ($Y_r$,$Z_r$)=($Y_{\rMS},Z_{\rMS}$). 
The observed split between the rMS and bMS is reproduced for 
$\DYDZ \simeq 200$, implying $\Delta Y = 0.144$ and $Y_{\bMS} = 0.39$. 
These values for $\Delta Y$ and $Y_{\bMS}$ are in excellent agreement 
with the results obtained by Piotto \etal (2005), Lee \etal (2005) 
and Sollima \etal (2007) 
by means of detailed isochrone analysis --- while the value for $\DYDZ$ may
vary significantly, depending on the assumed difference in metallicity
$\Delta Z$ in various studies. 
For instance, by adopting $Z_{\rMS}=0.001$ and $Z_{\bMS}=0.002$ 
as in Piotto \etal (2005), homology relations reproduce the observed 
temperature split with $\DYDZ \simeq 150$, implying again 
$\Delta Y \simeq 0.15$ as a robust result.
A rigorous transformation from $\mh$ to $Z$ (e.g.\ Casagrande \etal 2007) 
should take into account also the very different helium fraction in the two 
sequences ($Y_{\rMS}=0.246$ and $Y_{\bMS}=0.39$), returning $Z_{\rMS} = 0.00076$ 
and $Z_{\bMS} = 0.00120$; homology relations then yield $\DYDZ \simeq 320$, 
corresponding to $\Delta Y = 0.14$. Thus, $\Delta Y = Y_{\bMS}-Y_{\rMS}$ is a 
robust result of the analysis.

It is worth underlining that, in any case, in $\omega$Cen $\DYDZ >>70$, 
this value being correctly indicated by Piotto \etal (2005) as a 
{\it lower limit}, but often quoted in the literature as the preferred $\DYDZ$. 
Values of $\DYDZ \sim 200$ or higher are implied by the current metallicity 
measurements of the rMS and bMS (as also in Sollima \etal 2007), which 
renders the theoretical interpretation even more troublesome (see e.g.\ 
Yi 2009).

%%%%%%%%%%%%%%%%%%%%%%%%%%%%%%%%%%%%%%%%%%%%%%%%%%%%%%%%%%%%%%%%%%%%%%%%%%%%%%%
\subsection{NGC~2808}
Another striking example of helium enriched multiple Main Sequences is
NGC~2808, where three MSs are found at virtually the same metallicity 
$\feh =-1.1 \pm 0.03$~dex (Carretta \etal 2006; see also Salaris \etal 2006) 
and helium abundances of $Y=0.248$ (the value assumed for 
the red MS, which includes the bulk of the population), $Y=0.30$ (middle MS) 
and $Y=0.37$ (blue MS; D'Antona \etal 2005; Lee \etal 2005; Piotto \etal 2007).
In this case, formally $\DYDZ \rightarrow \infty$ (or, $\DYDZ > 800$) 
and one should just focus on $\Delta Y$. Also in this case, we show that
homology relations yield an estimate of $\Delta Y$ in good agreement 
with isochrone analysis.

As for wCen, the scale of Casagrande et~al.\ (2006) was used to convert
the colour split into a temperature split. In NGC2808, the maximum split
of the bMS and rMS with respect to the mMS ($\pm 0.08$~mag) occurs around
a magnitude $m_{\rm{F814W}}=21.3$, where the colour of the mMS is 
$\rm{F475W}-\rm{F814W} = 1.72$. Correcting for reddening 
(E(B-V)=0.18, Piotto et~al.\ 2007) 
and transforming the WFC/ACS Vegamag system photometry into Johnson--Cousins
according to Sirianni et al.\ (2005), these colours and broadening
correspond to $B-I = 1.74$ and $\Delta (B-I) \pm 0.11$~mag, equivalent to 
$T_{eff}=5020$~K for the mid MS and $\Delta \log \teff = \pm 0.01$.

As $\Delta Z$ vanishes and $\DYDZ$ diverges in the case of NGC~2808, 
it is convenient to recast the homology relation in Eq.~\ref{eq:DlogTeff} 
replacing $\delta \, (Z-Z_r) \rightarrow (Y-Y_r)+(Z-Z_r)$. 
Fig.~\ref{fig:ngc2808_homology} compares the predicted temperature split as 
a function of $\Delta Y$ to the observed one. 
For an assumed $Y_{\rMS}=0.248$, the resulting $Y_{\mMS}=0.32$ and 
$Y_{\bMS}=0.39$ are again in excellent agreement 
with isochrone analysis (see Fig.~2 of Piotto \etal 2007).

\subsection{47 Tuc}
Theoretical homology relations agree very closely with detailed 
isochrone analysis in the estimate of the $\Delta Y$ and $\DYDZ$,
characterizing the split of the multiple MSs of $\omega$Cen and NGC~2808.
We have checked that this holds true 
also for the (less extreme) case of 47 Tuc. The observed broadening of the MS
of 47 Tuc is $\Delta (F606W-F814W) = 0.013$ around an average colour 
$F606W-F814W \simeq 0.9$ at $F606W=20.5$; if interpreted
in terms of a dispersion in helium abundance, this implies $\Delta Y =0.026$
(Anderson \etal 2009). To analyze the same dispersion with homology relations,
we have translated the observed broadening into Johnson colours, and then into
effective temperatures. 

With the reddening and distance modulus listed in the catalogue by Harris 
(1996; in its 2003 online version) the broadening amounts to 
$\Delta (V-I)=0.017$ around $M_V = 7.3$ and $V-I = 1.04$.
With a metallicity $\feh = -0.76$ (Harris 2003; Carretta \etal 2009), this
average $V-I$ colour corresponds to $\teff \sim 4800$~K and the broadening
to $\Delta \log \teff=0.003$. Homology relations (Eq.~\ref{eq:DlogTeff})
reproduce such broadening with $\Delta Y=0.023$, again very close 
to the conclusions from isochrone fitting.

%%%%%%%%%%%%%%%%%%%%%%%%%%%%%%%%%%%%%
\begin{figure*}
\begin{center}
\includegraphics[scale=0.47]{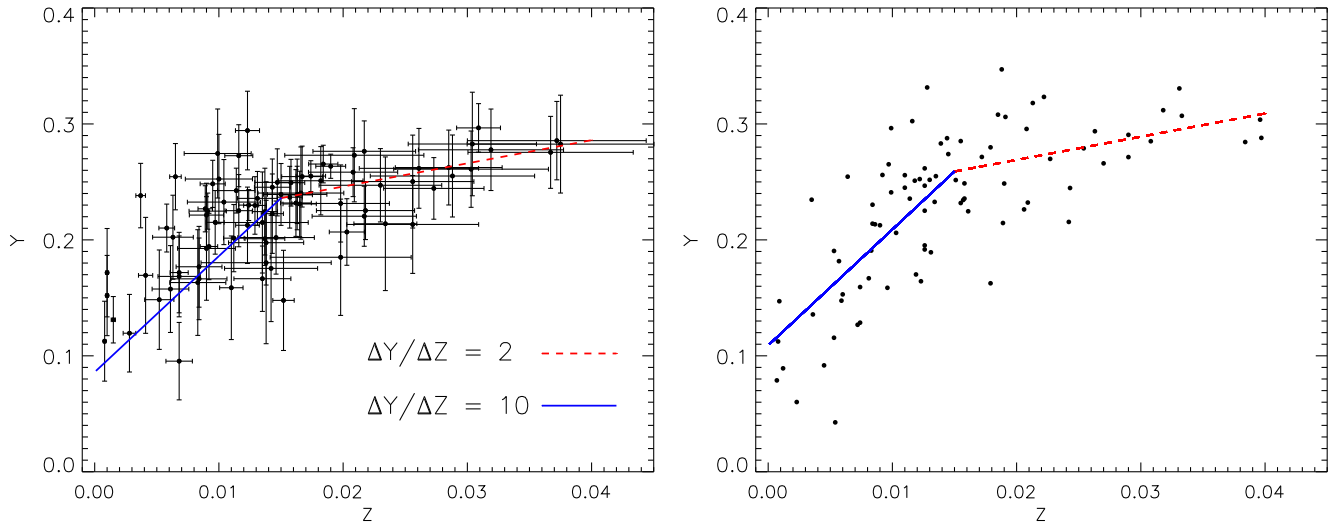}
\caption{Metal versus helium mass fraction for nearby field dwarfs. Lines with 
two values of helium--to--metal enrichment ratio (break at $Z=0.015$) are 
plotted to guide the eye. 
{\it Left panel:} using the isochrone fitting procedure described in 
Casagrande \etal (2007). {\it Right panel:} using numerical homology relations 
(see Section~\protect{\ref{sec:homology_num}}).}
\label{fig:YZfield}
\end{center}
\end{figure*}
%%%%%%%%%%%%%%%%%%%%%%%%%%%%%%%%%%%%%%%

%%%%%%%%%%%%%%%%%%%%%%%%%%%%%%%%%%%%%%%%%%%%%%%%%%%%%%%%%%%%%%%%%%%%%%%%%%%%%%%
\section{Homology relations and isochrones}\label{sec:homology_num}

We have shown that theoretical homology relations, applied to the multiple
(or broadened) MSs of globular clusters, provide a similar interpretation 
of the data in terms of $\Delta Y$ and $\DYDZ$, as detailed isochrone analysis.
We now compare directly homology relations 
to current stellar models. We fit the Padova isochrones as a 
function of $Z$ and $\DYDZ$ with a homology--like relation and compare the 
fitted relation to the theoretical one. 

An interpolation formula for isochrones directly inspired by 
Eq.~\ref{eq:DlogTeff} involves at least 4 independent 
parameters: the coefficients of the first 3 terms in Eq.~\ref{eq:DlogTeff},
and the coefficient of $Z$ within the argument of the logarithm 
in the $4^{th}$ term.
For fitting purposes, however, we found that a simplified form of the homology
relations is more convenient. We derive it by condensing together 
the first three terms in Eq.~\ref{eq:DlogTeff}, 
containing the dependence of the broadening on $\Delta Y/\Delta Z$. First
of all we notice that those three terms have a similar dependence:
\[ \propto \log \left[ 1-\frac{\delta}{(a+X_r)} (Z-Z_r) \right]~~~~~~~~~~~~~~~
a=0,~0.6~{\rm or}~1\]
as in the second term we can approximate:
\[ \frac{5 \delta +1}{(3+5 X_r-Z_r)}(Z-Z_r) \simeq 
\frac{5 \delta}{(3+5 X_r)}(Z-Z_r) \]
Secondly, it is easy to verify that the second term of 
Eq.~\ref{eq:DlogTeff} 
is the leading one, contributing about as much as the sum of the first 
and third term together.
Guided by these considerations, we find that the following formula:
\begin{equation} 
\label{eq:DlogTeff_2terms}
\begin{array}{l l}

\medskip
\Delta \log \teff = & -0.414 \log \left[ 1-\frac{\delta}{(0.6+X_r)}(Z-Z_r) 
                       \right] \\ 

 & -0.051 \log \left( \frac{100 Z+1}{100 Z_r+1}  \right) \\
\end{array}
\end{equation}
is a very good approximation of the rigorous homology relations, especially
for metallicities $Z \leq Z_{\odot}$ which are of interest for our discussion
(Fig.~\ref{fig:homology}). We have also verified that this is a fully adequate
approximation of the full homology relations 
even for extremely high $\Delta Y/\Delta Z > 100$, which 
are of interest for globular clusters, at least within narrow metallicity 
ranges (like those separating the sub-populations of e.g.\ $\omega$Cen).
This simplified formulation has the advantage of 
separating one term sensitive to the helium content, from the second term
which depends only on metallicity. This clearcut separation will prove to be 
handy for the empirical re-calibration of homology relations in Section~5.

Therefore we seek a three--parameter fitting formula for isochrones 
(and, later below, for real stars) of the kind: 
\begin{equation} 
\label{eq:DlogTeff_param}
\begin{array}{l l}
\medskip
\Delta \log \teff = & -P_1 \log \left[ 1-\frac{\delta}{(0.6+X_r)}(Z-Z_r) 
                       \right] \\

                     & -P_2 \log \left( \frac{P_3 Z+1}{P_3 Z_r+1}  \right) \\
\end{array}
\end{equation}
where $P_i$ are 3 free fitting parameters. It is our experience that this 
3-parameter formula provides an adequate fit to isochrones, as good as (and
more robust than) other homology-like fitting formul\ae\ with 4 or more
parameters. We favour this form of the homology 
relation over the linearized one adopted by Pagel \& Portinari (1998;
their eq.~5) as it is better suited to handle large values of $\DYDZ$ 
(see the Appendix).

We consider both the Padova isochrones computed specifically for the analysis 
in Casagrande \etal (2007), and the more recent release by Bertelli \etal 
(2008) with varying $\DYDZ$ and $0.0001 \le Z \le 0.04$, which fully brackets
the metallicity range relevant for the present study. 
In Casagrande \etal (2007) we had checked that other sets of isochrones 
(Yonsei--Yale, Teramo, McDonald) were very similar in the low MS.

We tried the fit both with respect to the solar ($Z_{\odot}=0.017$) and to a 
very metal-poor reference isochrone ($Z=0.0002$). 
We explored values of $\DYDZ$ up to 1000, obtained both
with sub-solar and sub-primordial helium contents 
(isochrones from Casagrande \etal 2007) and with helium--enhanced 
isochrones (Bertelli \etal 2008).
Very similar parameters are obtained in both cases, as one would expect 
from the discussion in Section~\ref{sec:homology}.

We find that the behaviour of isochrones, at least in the 
range $M_{\rm{Bol}}=5.4-7.0$ which is where the effect of $\DYDZ$ is expected 
to be maximal and dominant, is well described by homology--like relations 
with the 
following set of parameters:
\[ P_1=0.50  \pm 0.03 \]
\[ P_2=0.064 \pm 0.005\]
\[ P_3=670   \pm 200  \]
where the values are the averages from the fits obtained at three fixed 
$M_{\rm{Bol}}=5.4,6.0,6.5$ and $Z=0.0002$ and $0.017$. 
In the following of the discussion, when using these parameters we will 
speak of ``numerical homology relations''. With these parameters, 
the results obtained for field stars by Casagrande \etal (2007) are well 
reproduced (see Fig \ref{fig:YZfield}). In Casagrande \etal 
(2007) the metal and helium mass fraction of the stars were computed 
iteratively, with a star--by--star isochrone fitting, whereas now $Z$ is 
scaled with $Z_{\odot}$ and $Y$ computed via Eq.~\ref{eq:DlogTeff_param}. 
Despite the less sophisticated approach provided by homology relations, 
the overall agreement is good and the change in slope above and below 
$Z \sim 0.015$ is still reproduced. This confirms the validity of our 
simplified approach to study the broadening of the low MS. 

We notice that the fitting parameters are quite close to those of the 
(simplified) homology relation in Eq.~\ref{eq:DlogTeff_2terms}, with 
the exception of $P_3$ which is about 7 times larger. This was also noticed 
by Pagel \& Portinari (1998); the corresponding parameter in their formalism 
was $P_3=1/Z_0$, with $Z_0=0.01$ from the theoretical homology relations 
and $Z_0=0.0015$ from isochrone fitting. 
The discrepancy was mostly imputed to a difference between the observational 
and the theoretical HR diagram (Castellani \etal 1999) while here we confirm it
also in the purely theoretical plane.
 
The parameter $P_3$ describes the metallicity dependence 
of the opacity coefficient $\kappa_0$ (see the Appendix) so it is no surprise 
that it may differ between detailed stellar structure models and 
simplified descriptions of $\kappa_0$.
Opacity throughout the star (at least for largely radiative structures 
like low mass MS stars, where the convective envelope is extended in radius
but not so prominent in mass, and most importantly, energy production
occurs in radiative regions) is the main factor determining its luminosity 
and henceforth its whole structure; 
so a significant difference in the parameter $P_3$ implies a significant 
difference in structure (or, better to say, in its metallicity dependence) 
between actual stellar
models and homologous stars. However, $P_3$ is unrelated to $Y$ or $\DYDZ$ and,
since the other parameters ($P_1$ and $P_2$) are similar to the expected 
values, in essence homology relations render very well the dependence of 
theoretical isochrones on variations of the helium content
--- as our experiment on $\omega$Cen and NGC~2808 had already suggested.
%%%%%%%%%%%%%%%%%%%%%%%%%%%%%%%%%%%%%%%%%%%%%%%%%%%%%%%%%%%%%%%%%%%%%%%%%%%%%%%

\section{Empirical homology relations}

In the previous sections we have shown that theoretical homology relations
agree very well with the isochrone--based interpretation of the MS splits 
observed in $\omega$Cen and NGC~2808, and adequately reproduce the behaviour 
of isochrones in general, as a function of $\DYDZ$. However, isochrones are 
known to fail the interpretation of the HR diagram of nearby metal--poor
low MS stars 
(e.g.\ Lebreton \etal 1999; Torres \etal 2002; Casagrande \etal 2007;
Boyajian \etal 2008). Therefore, in this section we define ``empirical'' 
homology relations, calibrated to reproduce the broadening of the local MS, 
and extrapolate the consequences for the interpretation of globular clusters.

Casagrande \etal (2007) have shown that, while for 
$Z \gtrsim0.015$ the broadening of the main sequence is well reproduced
with the standard $\DYDZ \sim 2$, with current isochrones a much higher 
$\DYDZ \sim 10$ is needed to fit lower metallicities (Fig.~\ref{fig:YZfield}). 
While in principle $\DYDZ$ may not necessarily be constant,
Fig.~\ref{fig:YZfield}, taken at face 
value, also implies a helium content in local metal poor stars 
as low as $Y = 0.1$. Such a striking contrast with BBN is to be ascribed 
to inadequacies of low metallicity stellar models. 

While awaiting for a solution to this problem by improved stellar physics
(see our final discussion), here we adopt a very pragmatic approach: 
we {\it assume} that $\DYDZ \sim 2$ also for any $Z < Z_{\odot}$, and
empirically calibrate homology relations so that this result is returned
for nearby stars. Our assumption is very reasonable, since HII region 
measurements and chemical evolution models support a constant $\DYDZ \sim 2$ 
down to very low $Z$ (e.g.\ Peimbert \etal 2007; 
Carigi \& Peimbert 2008), 
as does the following simple argument: taking the solar bulk 
abundances (Asplund \etal 2009) and the primordial $Y_P=0.240$ (Steigman 2007), 
one derives $\DYDZ = 2.1$ for $Z<Z_{\odot}$.

To define our empirical homology relations, let's first inspect the form
of the relations (Eq.~\ref{eq:DlogTeff_param}). The first term includes the
effects of the helium content, while the second term (stemming essentially
from the metallicity dependence of opacity) is independent of $\DYDZ$. 
The second term is therefore irrelevant for the 
multiple MSs in GCs, where metallicity differences are small or vanishing. 
Therefore, if we calibrate our empirical homology relations onto nearby stars 
acting only on the first term (via the parameter $P_1$), we maximize the change
in the role of helium, and hence maximize the consequences for the 
interpretation of GC's.\footnote{Ideally, one would use the data to calibrate 
all of the three parameters at the same time, or even to calibrate more complex
relations such as Eq.~\ref{eq:DlogTeff}, which would yield some handle on the 
corresponding physical ingredients. Unfortunately, the data is too noisy 
to allow for such a detailed calibration.}

We shall therefore start by discussing how the first term of
Eq.~\ref{eq:DlogTeff_param} must change, to yield an acceptable
$\DYDZ \sim 2$.
This correspond to imputing the erroneous theoretical broadening 
of the main sequence entirely to the response of stellar structure to the 
helium--to--metal enrichment ratio $\DYDZ$.

%%%%%%%%%%%%%%%%%%%%%%%%%%%%%%%%%%%%%%%%%%%%%%%% 
\begin{figure}
\begin{center}
\includegraphics[scale=0.85]{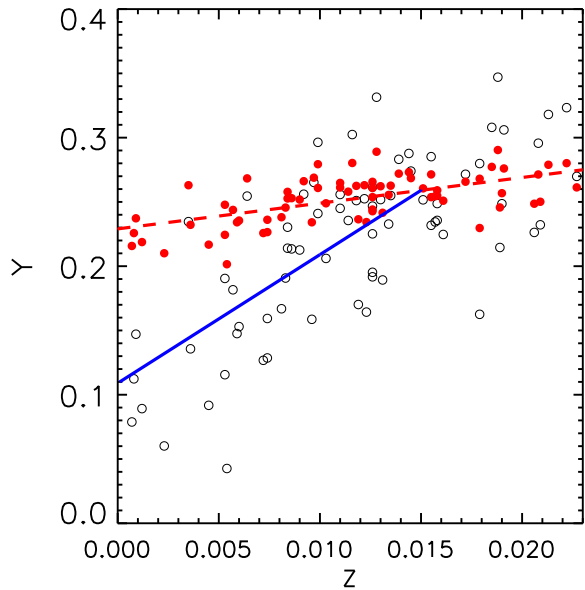}
\caption{Metal versus helium mass fraction for the dwarfs in the right panel 
of Figure \ref{fig:YZfield} (zooming on metallicities around and below solar).
Open circles are using the numerical homology relations established 
in Section~\ref{sec:homology} from isochrone fitting, 
while full circles are obtained using $P_1=1.5$ as discussed in the text. 
A line with constant $\DYDZ=2$ is also shown for comparison.}
\label{fig:YZpoor}
\end{center}
\end{figure}
%%%%%%%%%%%%%%%%%%%%%%%%%%%%%%%%%%%%%%%%%%%%%%%

%%%%%%%%%%%%%%%%%%%%%%%%%%%%%%%%%%%%%%%%%%%%%%%%%%%%%%%%%%%%%%%%%%%%%%%%%%%%%%%
\subsection{Empirical calibration of $P_1$ and globular clusters}

Extant isochrones, or the equivalent ``numerical'' homology relations
(Eq.~\ref{eq:DlogTeff_param} with $P_1=0.50$, $P_2=0.064$ and $P_3=670$)
yield, for the observed broadening $\Delta \log \teff$ at low $Z$, a resulting
$\delta= 1 + \DYDZ \sim 11$. As we aim at imposing $\delta=3$, we can expect
that we will need to increase $P_1$ by a factor of the order of 11/3, i.e.\
$P_0 = 0.50 \longrightarrow \sim 2$. This is easily seen e.g.\ with 
a Taylor expansion of the logarithm, so that the first term approximately
goes as $ \sim P_1 \frac{\delta}{(0.6+X_r)}(Z-Z_r)$: for a given 
$\Delta \log \teff (Z)$, $P_1$ and $\delta$ are inversely proportional,
and an increase of the former by 11/3 corresponds to a reduction of the latter 
by the same amount.

More rigorously, we keep $P_2=0.064$ and $P_3=670$ fixed and optimize
$P_1$ so that the inferred helium content of nearby K dwarfs with $Z<Z_{\odot}$
follows $\DYDZ =2$. This optimization yields $P_1 =1.5$ (similar to our
simple expectations above) and the helium abundances shown in 
Fig.~\ref{fig:YZpoor}.

%%%%%%%%%%%%%%%%%%%%%%%%%%%%%%%%%%%%%%%%%%%%%%%%%%%%%%
\begin{figure*}
\begin{center}
\includegraphics[scale=0.43]{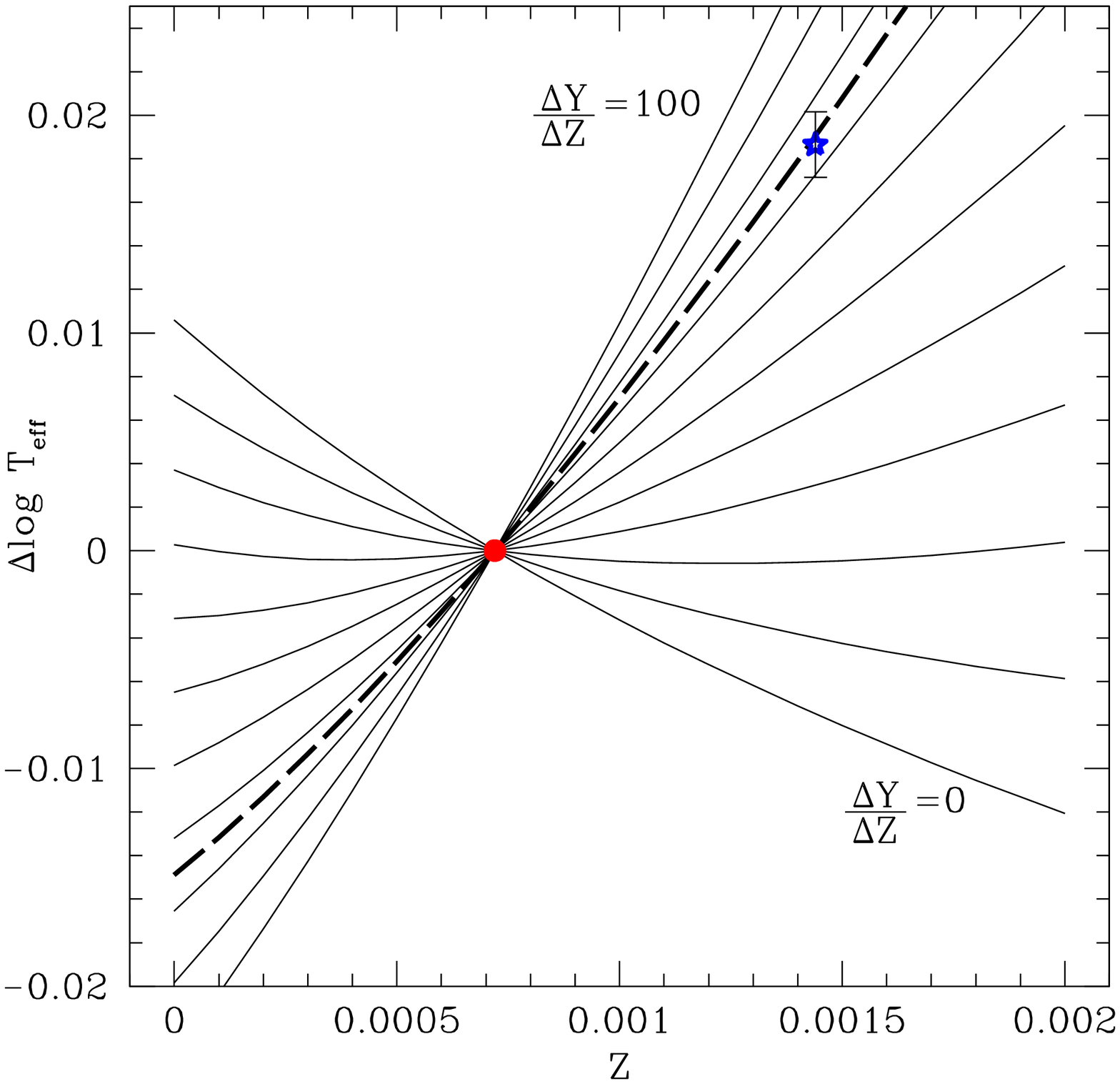}
\includegraphics[scale=0.43]{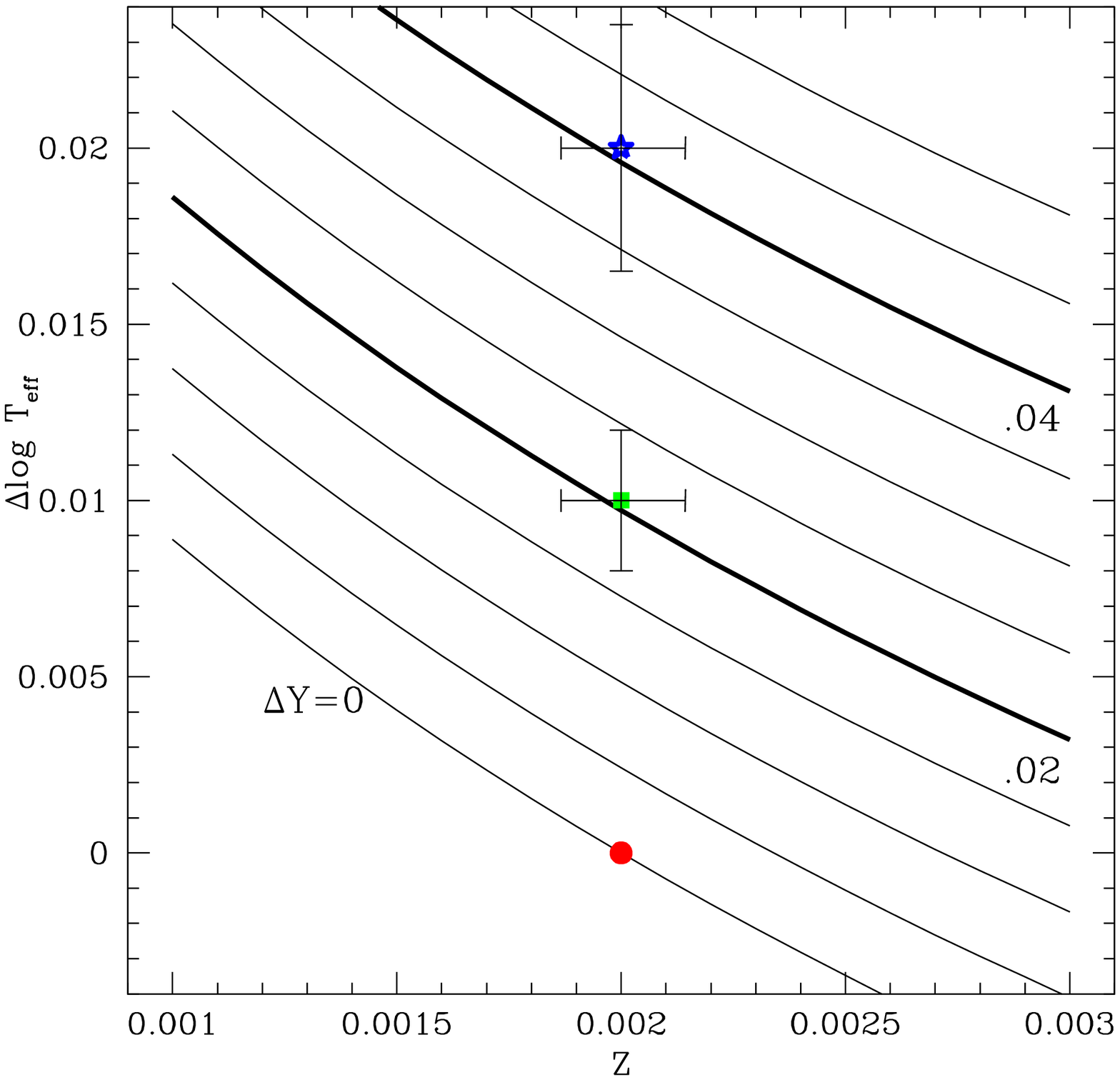}
\caption{{\it Left (right) panel:} same as Figure \ref{fig:WCen_homology} 
(\ref{fig:ngc2808_homology}) for $\omega$Cen (NGC~2808), but using Eq.\ 
\ref{eq:DlogTeff_param} with $P_1=1.5$.}
\label{fig:newhomology}
\end{center}
\end{figure*}
%%%%%%%%%%%%%%%%%%%%%%%%%%%%%%%%%%%%%%%%%%%%%%%%%%%%%%%

We now re--interpret the split of the multiple MS of globular clusters
by means of these empirical homology relations, with a re-calibrated
$P_1=1.5$. Fig.~\ref{fig:newhomology} shows that the blue sequence of 
$\omega$Cen is now fitted with $\DYDZ=75$ and $Y_{\bMS}=0.30$.
For NGC~2808, the middle and blue MS are now fitted with $\Delta Y=0.02$
and $0.04$ respectively, corresponding to $Y_{\mMS}=0.27$ and $Y_{\bMS}=0.29$.
These new values of the helium enrichment are significantly lower than
the earlier estimates $Y \simeq=0.4$ and are within reasonable reach
of extant theories on stellar nucleosynthesis and chemical evolution
(e.g.\ Yi 2009). 

To summarize: if we assume that the unique culprit of the erroneous broadening 
of the theoretical low--Z MS is the response of stellar structure to 
$\DYDZ$, and consequently recalibrate the first term in the homology relations,
the estimated $\Delta Y$ for the subpopulations of GC's is drastically
reduced. The new estimates of the helium content of the subpopulations
are no longer extreme; rather, they are close to the solar helium content 
or slightly larger ($Y \leq 0.3$) which makes them relatively easy to reconcile
with chemical evolution models.\footnote{This new estimate was obtained 
by recalibrating $P_1$ on the low--$Z$ range alone, i.e.\ using observed
K dwarfs with $Z<Z_{\odot}$ (Fig.~\ref{fig:YZpoor}), 
as for $Z>Z_{\odot}$ isochrones yield the correct
$\DYDZ \sim 2$ as they are. If we rather want to calibrate the homology
relations over the whole metallicity range of the sample, so that the same
homology formula yields $\DYDZ \sim 2$ between $0<Z<0.04$, we get a lower
$P_1=1.1$ as the optimized value. The interpretation
of GCs remain similar: for $\omega$Cen, $\DYDZ=100$ and $Y_{\bMS}=0.32$;
for NGC~2808, $Y_{\mMS}=0.28$ and $Y_{\bMS}=0.30$.}

%%%%%%%%%%%%%%%%%%%%%%%%%%%%%%%%%%%%%%%%%%%%%%%%%%%%%%%%%%%%%%%%%%%%%%%%%%%%%%%%
\subsection{Empirical calibration of the second term}

An alternative way to bring the theoretical broadening of the low MS 
in agreement with observations, is to act on the second term of the homology
relations in Eq.~\ref{eq:DlogTeff_param}. As anticipated above, this
term only depends on metallicity and therefore its alteration will have
no impact on the interpretation of the multiple MS of globular clusters,
where metallicity differences are minimal or vanishing. But for nearby
K dwarfs and subdwarfs, the metallicity range is significant and exploring
the effects of this second term is worthwhile. 

\begin{figure*}
\begin{center}
\includegraphics[scale=0.7]{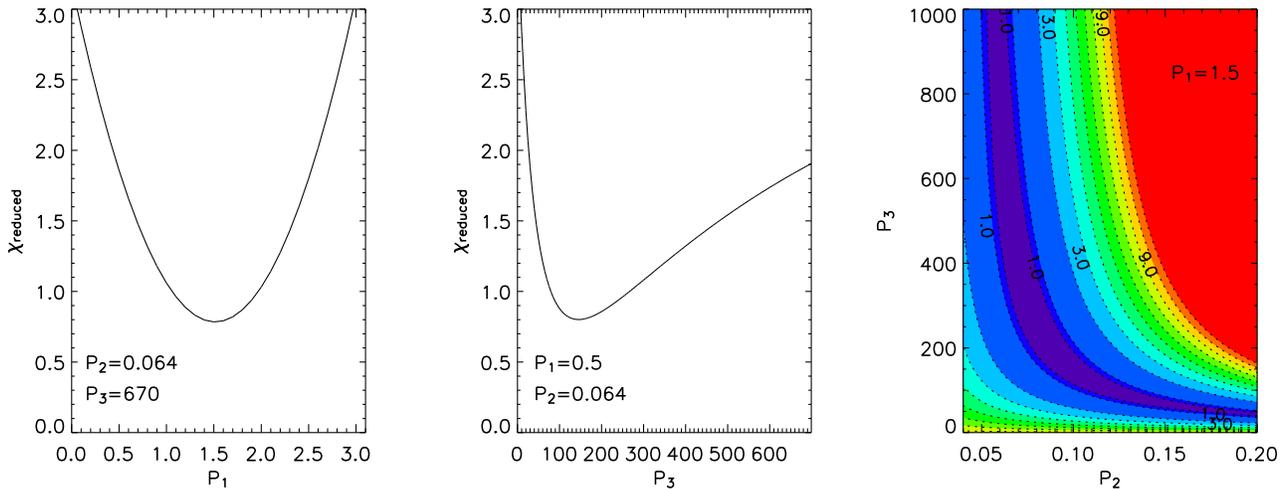}
\caption{{\it Left and central panel:} reduced $\chi^{2}$ obtained fitting the 
K dwarfs in Casagrande \etal (2007) to Eq.\ \ref{eq:DlogTeff_param},
imposing $\DYDZ=2$ and optimizing only one free parameter at a time.
{\it Right panel:} contour 
lines for the reduced $\chi^2$ (values indicated by the labels) when 
only $P_1$ is fixed: the best solution for ($P_2$, $P_2$) lies along an 
hyperbola.}
\label{fig:contour}
\end{center}
\end{figure*}

It is clear from (Eq.~\ref{eq:DlogTeff_param}) that parameters $P_2$ and
$P_3$ are degenerate, in the sense that a given change in the second term
can be obtained by modifying either of the two parameters.
Fig.~\ref{fig:contour} shows that, fixing $P_1$, the best combined solution for
$P_2$ and $P_3$ lies roughly along a hyperbola.

Therefore, we choose to discuss the role of the second term
by optimizing $P_3$, which is an interesting parameter as it is 
significantly different between the theoretical and the numerical 
homology relations (Section~\ref{sec:homology_num}). 

\begin{figure}
\begin{center}
\includegraphics[scale=0.85]{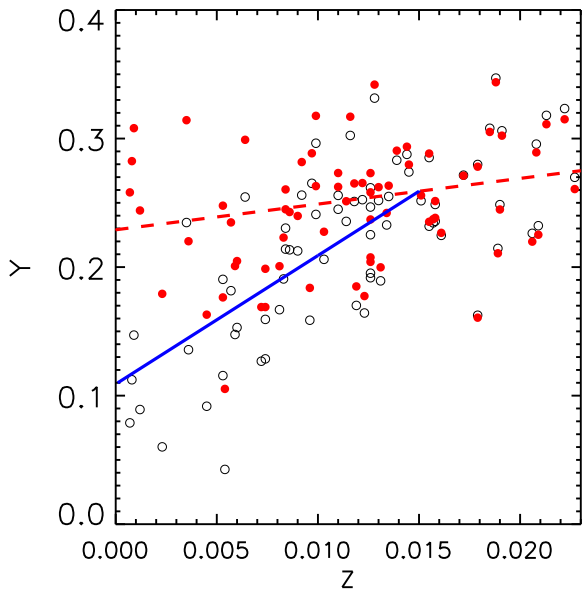}
\caption{Same as Figure \ref{fig:YZpoor} but when operating on $P_3=150$ to 
obtain the $\DYDZ=2$.}
\label{fig:YZ_P3}
\end{center}
\end{figure}

Keeping $P_1$ and $P_2$ fixed to the values of 
Section~\ref{sec:homology_num},
we find that $P_3 \sim 150$ is the optimal value to obtain $\DYDZ=2$
for nearby low--Z stars (Fig.~\ref{fig:YZ_P3}). However, while we do obtain 
$\DYDZ=2$ on average, compared to the case in Fig.~\ref{fig:YZpoor}
there is now considerable scatter and many stars remain with uncomfortably low
helium abundances. It is intriguing, though, that the optimized value of
$P_3$ is quite close to the theoretical value of 100.

%%%%%%%%%%%%%%%%%%%%%%%%%%%%%%%%%%%%%%%%%%%%%%%%%%%%%%%%%%%%%%%%%%%%%%%%%%%%%%%
%%%%%%%%%%%%%%%%%%%%%%%%%%%%%%%%%%%%%%%%%%%%%%%%%%%%%%%%%%%%%%%%%%%%%%%%%%%%%%%

%%%%%%%%%%%%%%%%%%%%%%%%%%%%%%%%%%%%%%%%%%%%%%%%%%%%%%%%%%%%%%%%%%%%%%%%%%%%%%%
%%%%%%%%%%%%%%%%%%%%%%%%%%%%%%%%%%%%%%%%%%%%%%%%%%%%%%%%%%%%%%%%%%%%%%%%%%%%%%%

Notice that the cause of the discrepancy is entirely ascribed to 
some sort of metallicity dependence now. For instance, the variation 
of the mixing-length with $Z$ suggested in Casagrande \etal (2007) 
to avoid low helium abundances, 
in the framework of the homology relations can be formally 
described by the second term. The same can be said for any change in stellar 
models that would only respond to metallicity differences and not to 
differences in the helium content:
it would have no impact on the interpretation of the multiple main sequences 
of Globular Clusters.

%%%%%%%%%%%%%%%%%%%%%%%%%%%%%%%%%%%%%%%%%%%%%%%%%%%%%%%%%%%%%%%%%%%%%%%%%%%%%%%
\section{Summary and discussion}

In this paper we draw attention on the possible connection between
the HR diagram of globular clusters with multiple Main Sequences
($\omega$Cen and NGC~2808) and the broadening of the low MS defined
by nearby stars.
Although not ``inverted'' as in $\omega$Cen, where the metal-rich Main 
Sequence is bluer than the metal-poor one, the broadening defined by nearby 
subdwarfs is much narrower than expected for a standard helium content
and helium--to--metal enrichment law $\DYDZ \sim 2$.
At low metallicities, $\DYDZ\sim 10$ is formally necessary to reproduce 
the observed broadening, but this is unacceptable as it implies
helium fractions much lower than the cosmological BBN floor. 

It is worth remarking here that the results by Casagrande \etal (2007) on the
broadening of the low MS  were corroborated by the analysis of a number of
binary stars, whose masses were well reproduced by theoretical isochrones --- 
again at the expense of assuming, for some low Z binaries, sub-primordial
helium abundances.
Also, one may argue that the problem lies in an incorrect $\teff$
scale for low Z stars. Without entering here a detailed discussion on the
robustness of the IRFM scale by Casagrande \etal (2006, 2010), suffice here
to mention that this is one of the hottest scales available 
(as discussed in the original papers and in Sousa et al.\ 2008).
It is hotter by 100~K than other IRFM renditions (the now superseded scales 
by Alonso, Arribas \& Martinez--Roger 1996 and Ram\'irez \& Mel\'endez 2005), 
and it is comparable within $\pm$50~K to various spectroscopic scales
in the metallicity range relevant for this work.

Therefore, any other $\teff$ scale will just worsen the K dwarf problem, 
with real stars even cooler than the models. Also, it does not seem
plausible that all the (independent) scales available are systematically 
offset by $> 200$~K, which is what is needed to bring the K dwarfs
of Casagrande et al.\ (2007) in line with stellar models.

We estimate the possible extent of the required revision of low MS stellar 
models on the base of homology
relations. First we show that theoretical homology relations properly
reproduce the response of stellar models to the helium content: in particular,
analyzing the split of the multiple MS in $\omega$Cen and NGC~2808 by homology
relations yields the same conclusions as full isochrone analysis (i.e.\
a helium content $Y \sim$ 0.4 for the blue subpopulations). Then,
since both isochrones and theoretical homology relations fail the 
interpretation of the nearby low--Z Main Sequence, we calibrate {\it empirical}
homology relations to yield consistently $\DYDZ\sim 2$
for nearby stars, and inspect the consequences for the MS splits in globular
clusters.

{\it If} we entirely impute the failure of the low-MS stellar models to 
a wrong response to the helium abundance (and correspondingly re-calibrate 
the helium--sensitive term of the homology relations), the consequences
for globular clusters are highly significant: the helium content of the blue
sub-populations is reduced from 40\% to 30\%, which is far easier to 
explain with chemical evolution models (Renzini 2008; Yi 2009).

Alternatively, if stellar models for the low MS are assumed to fail 
in their metallicity dependence (i.e.\ we re-calibrate only the homology term
expressing the metallicity dependence of opacity)
the consequences for globular clusters are negligible --- as
the metallicity differences between the subpopulations are minimal or vanishing.

As the solution for K dwarf models can be intermediate between these two
extreme assumptions, we suggest that the helium rich populations in globular 
clusters
are likely to have a helium content $Y$ in between 0.3--0.4; but altogether,
there is room to decrease their estimated helium content from the extreme 
$Y=0.4$ that is the commonly quoted value.

In summary, the purpose of this exercise is to draw the attention of 
model--makers to the problem of the HR diagram of nearby low--Z stars: 
improvements in this respect are potentially important also for the
riddle of the helium self-enrichment in globular clusters.

We remind indeed that, while a helium content as high as $Y \sim 0.35-0.4$ 
seems to nicely account for the multiple Main Sequences and the morphology
of the Horizontal Branch (HB), other observations provide some counter-evidence
for such a helium rich sub-population. 
In $\omega$Cen, the location of the
RGB bumps of the metal--poor and metal--intermediate populations is
consistent with a maximum $\Delta Y < 0.1$, less than what is derived 
from MS and HB analysis (Sollima \etal 2005). Sollima \etal (2006) have also
identified RR Lyr\ae\ stars with metallicity corresponding to that of the 
blue MS, but normal helium content --- which would require the 
metal--intermediate population to be further split into a helium rich and a
helium normal subcomponents.

NGC 6752 is another cluster suspected to host a helium enriched population,
due to the morphology of its HB and the presence of a broadened, possible
multiple MS (Milone et al.\ 2010); however, Villanova \etal (2009) did not 
detect
large helium abundances in the spectra of blue HB stars in the (narrow but 
crucial) temperature range $8500 < \teff < 11500$, hot enough to produce 
helium lines but still unaffected by helium sedimentation.
Other studies of clusters
suspected to host helium rich sub-populations have not confirmed their 
presence (e.g.\ Lee \etal 2009 for NGC 1851; Catelan \etal 2009b for M3). 
While awaiting for independent proof of high helium abundances
from direct spectral measurements or other features in the HR diagram
(see the recent review by Catelan \etal 2009a), we remark that the helium 
abundances
obtained by our proposed revision, $Y \lsim 0.3$, are consistent with
all of the above--mentioned constraints.

The need for sub-primordial helium abundances to fit a handful of nearby 
subdwarfs on the HR diagram was first noticed by Fernandes \etal (1998) and 
Lebreton \etal (1999) who advocated the inclusion of additional 
physical processes not implemented in standard stellar evolutionary 
calculations. In the following, we discuss a number of possible solutions
to the problem. Broadly speaking, we can classify them as 
metallicity--dependent
solutions, which shall not concern the helium rich populations of globular 
clusters, and helium--dependent solutions, which we expect also to impact
the interpretation of globular clusters. We may add that, considering the
nice agreement between the multiple MS and the HB morpology in $\omega$ Cen
and NGC~2808, the optimal helium--dependent solution should preferably
preserve this relation; this can be achieved if the new stellar models
will not just shift the low metallicity ZAMS to lower effective temperatures, 
but also affect the luminosity of individual stars and accelerate their 
evolution, which helps to populate the blue side of the HB.
Casagrande \etal (2007) discussed a number of possible solutions
(all in the metallicity--dependent class). We briefly recall them and discuss
a few more here below.

\paragraph*{Mixing length} 
A mixing length parameter decreasing with metallicity would make low Z models 
redder; for the isochrones we used, the variation should be from the reference 
solar value $\alpha$=1.68 to $\alpha$=1.0 at low Z.
While any dependence of $\alpha$ on metallicity,
mass or other physical parameters is still highly disputed (Casagrande \etal
2007 and references therein) this solution would have no consequence for the
helium content of globular clusters, as it is metallicity dependent and
does not significantly affect the luminosity and lifetime of the stellar models.
 
\paragraph*{Diffusion + non--LTE effects} 
Element diffusion affects the location of low MS stars in the HR diagram: 
it makes stellar models redder, and moreover it lowers the measured surface 
metallicity with respect to the real intrinsic one.
In this scenario, low Z subdwarfs look redder than we expect, mostly because 
they are actually more metal rich than we measure. 
Casagrande \etal (2007) discussed this possibility resorting to literature 
models with fully efficient (``maximal'') diffusion, and found that it does 
not completely solve
the problem, unless it is combined with additional errors on the observed
metallicity due to non--LTE effects. (This solution also would have no effect
on globular clusters, since it is metallicity and age dependent, and 
the gap in both is small for 
the multiple populations of globular clusters.)

However, as discussed in that paper, there is evidence that both of these 
mechanisms are not fully efficient in real stars, as one would need to solve 
the K dwarf problem. For diffusion in particular,
observations of field and cluster metal-poor
dwarfs point toward inhibited efficiency and additional
--- yet ad hoc --- processes are nowadays invoked to contrast diffusion
(e.g.\ Chaboyer \etal 2001; Richard, Michaud \& Richer 2002, 2005; 
Korn \etal 2007).

\paragraph*{Boundary conditions} 
also play a role in stellar evolutionary tracks.
A solar scaled $T-\tau$ relation provides a good description of the atmospheric
structure also in metal-poor stars, and may contribute to improve low Z models 
with respect to gray atmosphere boundary conditions. However, the effect 
appears to be quite small for MS stars, when each set of isochrones 
consistently adopts a mixing length parameter calibrated on the solar model
(Vandenberg \etal 2008). 

Besides, boundary conditions are expected to become more relevant 
at lower stellar masses and luminosities (due to the deeper
surface convection), while in Casagrande \etal (2007) we verified that 
there is no systematic trend of the derived $\DYDZ$ with $M_{bol}$.

Also this solution relies on a metallicity--dependent effect and would not 
bring substantial revisions to the helium content in the multiple MS of GCs.

\paragraph*{Convection} 
One can always consider changing the boundary of convective regions 
as a viable working hypothesis.
(In fact, the very presence of a convective envelope renders real K dwarfs
not rigorously representable by homology relations.) Considering that the
convective envelope gets thinner at lower metallicities, any change
in the convection scheme (e.g. extra-mixing, or undershooting) is expected
to affect solar metallicity objects more, creating a differential effect
in metallicity that could change the relative location of the low MS as 
a function of Z. Also this solution falls in the metallicity-dependent
category and is not expected to impact on the GC analysis.

\paragraph*{Opacity} 
As K dwarfs are largely radiative structures, in particular  
in the regions where nuclear energy is produced, opacity is a leading
ingredient in determining their luminosity. An incorrect metallicity dependence
of opacity would seriously affect the broadening of the low MS. The
idea is tempting, as the ``break'' in the estimated $\DYDZ$ occurs close to
$Z_0=0.01$, which is the classic divide between free-free and bound-free 
dominated opacity in stellar interiors (Cox \& Giuli 2004). The parameter $P_3$ 
in the original homology relations is related to this characteristic 
metallicity as $P_3=1/Z_0$ (Section~\ref{sec:homology_num} 
and Appendix) and we have seen that modern isochrones are better described
by $P_3 \simeq 670$ rather than 100, effectively reducing $Z_0$ to 0.0015
(Section~\ref{sec:homology_num}; Pagel \& Portinari 1998). However, 
optimizing empirical homology relations on the opacity term, restores 
a value for $P_3$ or $Z_{ff/bf}$ closer to the theoretical one ($P_3=150$ 
or $Z_0\simeq 0.007$). This suggests that the culprit might be
the metallicity  dependence of opacity (which would then be irrelevant for 
globular clusters). It is easier to figure errors in the bound-free contribution
to opacity, rather than in the free-free component, at least for the bulk
of the star wher H and He are completely ionized. Lowering the bound-free 
contribution to opacity would imply a recalibration of solar metallicity 
isochrones, after which high Z (bound-free dominated) and low Z (free-free 
dominated)
model MS may fall closer to each other, thus reducing the broadening. 
However, we wonder if there is much 
room for profound changes in opacity, given the excellent agreement
between the major current, independent databases (Opacity Project, Seaton 2005 
and references therein; and OPAL, Iglesias \& Rogers 1996 and references 
therein) and keeping in mind that improvement in atomic data, 
input physics etc. most often leads to an increase in the opacity, 
as more and more opacity sources are taken into account. 
Helioseismology has shown extant OPAL and OP opacities to be 
fully adequate for Solar models with the ``classic'' solar composition; while 
an increase in opacity by 10--20\% in a suitable temperature range has been 
invoked for the Solar model with new, lower metallicity --- possibly 
in connection with increased neon abundance (Basu \& Antia 2008; 
Asplund et~al.\ 2009; Serenelli et~al.\ 2009; and references therein). How these
or other fundamental changes of the Solar model, to recover agreement with
helioseismological constraints, will impact low Z 
subdwarfs and the {\it relative} location of MS of subsolar metallicity
on the HR diagram, remains to be explored.

\medskip \noindent
Finally, we notice that the ``helium problem'' may not be limited 
to K dwarfs: systematic temperature offsets from the theoretical Main Sequence 
at low Z have been highlighted
for FG dwarfs in the Geneva--Copenhagen survey (N\"ordstrom et~al.\ 2004); 
and a helium content Y=0.23--0.24 (i.e.\ slightly sub-cosmological) has been
recently suggested for a slighly metal-poor F dwarf binary ([Fe/H]=--0.25; 
Clausen et~al.\ 2010). If the problem indeed extends to FG dwarfs, clearly
some of the suggested solutions (for instance, those related to diffusion)
are not viable.

\medskip \noindent
All of the solutions suggested above rely, more or less 
indirectly, on metallicity dependence; this is the standard way we think
of stellar models. However, our exercise in this paper highlights that it is
worth thinking of other possible systematics in the stellar models, especially
connected to the helium abundance, for their interesting impact on the 
multiple MS of globular clusters.

As mentioned above, opacity has a key role in the structure of lower MS stars,
so we may wonder whether the issue can be the helium dependence 
of opacity. This can be hardly modified in regions of complete ionization;
more promising are regions where He is partially recombined and contributes 
to the boud-free opacity.

Other helium--dependent effects may be considered, such as diffusion mechanisms
that would act differently for helium as for metals. (However, preliminary 
tests on only--helium diffusion seem to go in the opposite direction as needed
for globular clusters, with the helium rich MS getting proportionally 
more red than the helium poor; A.\ Serenelli, priv.\ comm.)

Possibly, other of the above mentioned solutions (mixing length, convection 
etc.) can be recast and explored in terms of helium dependence, rather than 
metallicity dependence. Any ideas in this direction are highly desired:
hopefully, astrophysicists will be as creative in solving the problem 
of the HR diagram of low metallicity K dwarfs, as they have been in tackling 
the riddle of the extreme helium rich populations in globular clusters!

\section*{Acknowledgments}
We acknowledge useful discussions with Achim Weiss and Aldo Serenelli.
We would like to remember here the late Prof. Bernard Pagel for his life-long 
interest in $\DYDZ$ and for introducing two of us (LP and CF) to the study
of K dwarfs as tracers of Galactic chemical evolution.
LP and CF acknowledge fundings from the Academy of Finland.

%%%%%%%%%%%%%%%%%%%%%%%%%%%%%%%%%%%%%%%%%%%%%%%%%%%%%%%%%%%%%%%%%%%%%%%%%%%%%%%%

%%%%%%%%%%%%%%%%%%%%%%%%%%%%%%%%%%%%%%%%%%%%%%%%%%%%%%%%%%%%%%%%%%%%%%%%%%%%%%%
\appendix
\section[]{Homology relations --- the background}
An excellent introduction of the homology relations used in this paper is given
by Fernandes \etal (1996), which we follow. A family of permanently homologous
stars (i.e., stars with the same relative mass distribution, that are 
in hydrodynamical and thermal equilibrium) can be obtained in the hypothesis 
that the equation of state is the perfect gas law, and that opacity and  
energy generation rate follow laws of the kind:
\[ \kappa = \kappa_0 \, \rho^n T^{-s} \]
\[ \epsilon = \epsilon_0 \, \rho^\lambda T^\nu \]
(Cox \& Giuli 1968). Other underlying assumptions are that the energy transport
occurs via radiative diffusion over most of the interior (in particular, over
the energy generation region, as is the case for the low MS) 
and that the radial profile of chemical composition differs, in different
stars, just by a scale factor.
The latter assumption holds in particular for stars on the ZAMS,
with internally uniform chemical composition.

For low--mass MS stars, opacity is dominated by bound--free and free-free 
processes and well approximated by Kramer's law: 
$\kappa = \kappa_0 \rho T^{-3.5}$;
while the energy generation rate via pp chain approximately follows
$\epsilon_{pp} =\epsilon_0 \rho T^4$. 
With these dependencies on $\rho$ and $T$, 
homology transformations yield the following relation for the luminosity
and effective temperature of the star
\[ L = \epsilon_0^{-0.077} \mu^{7.769} \kappa_0^{-1.077} M^{5.462} \]
\[ \teff = \epsilon_0^{-0.096} \mu^{2.211} \kappa_0^{-0.346} M^{1.327} \]
as a function of mass $M$ and molecular weight $\mu$.
Eliminating mass, one derives:
\begin{equation}
\label{eq:LTeff}
L = \epsilon_0^{0.318} \, \mu^{-1.332} \, \kappa_0^{0.347} \, \teff^{4.116}
\end{equation}
The last factor indicates the slope of the ZAMS for a given chemical 
composition: about 10 mag per dex in log($\teff$) --- flatter than, 
but not so far from, the slope obtained with detailed stellar structure 
computations (17 mag per dex, Section~\ref{sec:homology}).
Here we are mostly interested 
in the broadening of the ZAMS in luminosity as a function of chemical 
composition, at a fixed $\teff$; such broadening is expressed by the first 
three factors. 
Since H--burning occurs via the p-p chain:
\[ \epsilon_0 \propto X^2 \] 
The molecular weight $\mu$, throughout most of the star where complete 
ionization applies, is given by:
\[ \mu^{-1} = 2 X + \frac{3}{4} Y + \frac{1}{2} Z = 
             \frac{1}{4} ( 3 + 5 X - Z ) \] 
For bound--free and free--free opacity, approximately: 
$\kappa_{0,bf} \approx 7.40 \times 10^{24} \times \frac{5}{10} Z (1+X)$,
$\kappa_{0,ff} \approx 3.76 \times 10^{22} (1+X)$ (Cox \& Giuli 2004), so that
\[ \kappa_0=\kappa_{0,bf}+\kappa_{0,ff} \propto (1+X) (100 Z +1) \]
Notice that $Z_0 = \frac{1}{100}$ is where opacity drifts from free-free
dominated ($Z<0.01$) to bound-free dominated ($Z>0.01$).
Altogether, we can write the homology relation~(\ref{eq:LTeff}) as:
\[ L = {\cal F} (\epsilon_0,\mu,\kappa_0) \; \teff^{4.116} \]
with
\[ {\cal F} \propto X^{0.636} \, (3+5X-Z)^{1.332} \, 
(1+X)^{0.347} (100 Z+1)^{0.347} \]
The first factor in the $\cal F$ function is related to the energy generation
rate, the second factor to the molecular weight, and the last two factors
to opacity.
It is then straightforward to derive Eq.~\ref{eq:DMbol} for the broadening 
of the ZAMS in luminosity, with respect to a reference 
composition $(X_r, Y_r, Z_r)$:
\[ \Delta M_{\rm{Bol}} = -2.5 \log \frac{L}{L_r} = 
-2.5 \log \frac{{\cal F} (X,Y,Z)}{{\cal F} (X_r, Y_r, Z_r)} \]
% 
%%%%%%%%%%%%%%%%%%%%%%% Figure A1
\begin{figure*}
\begin{center}
\includegraphics[scale=0.43]{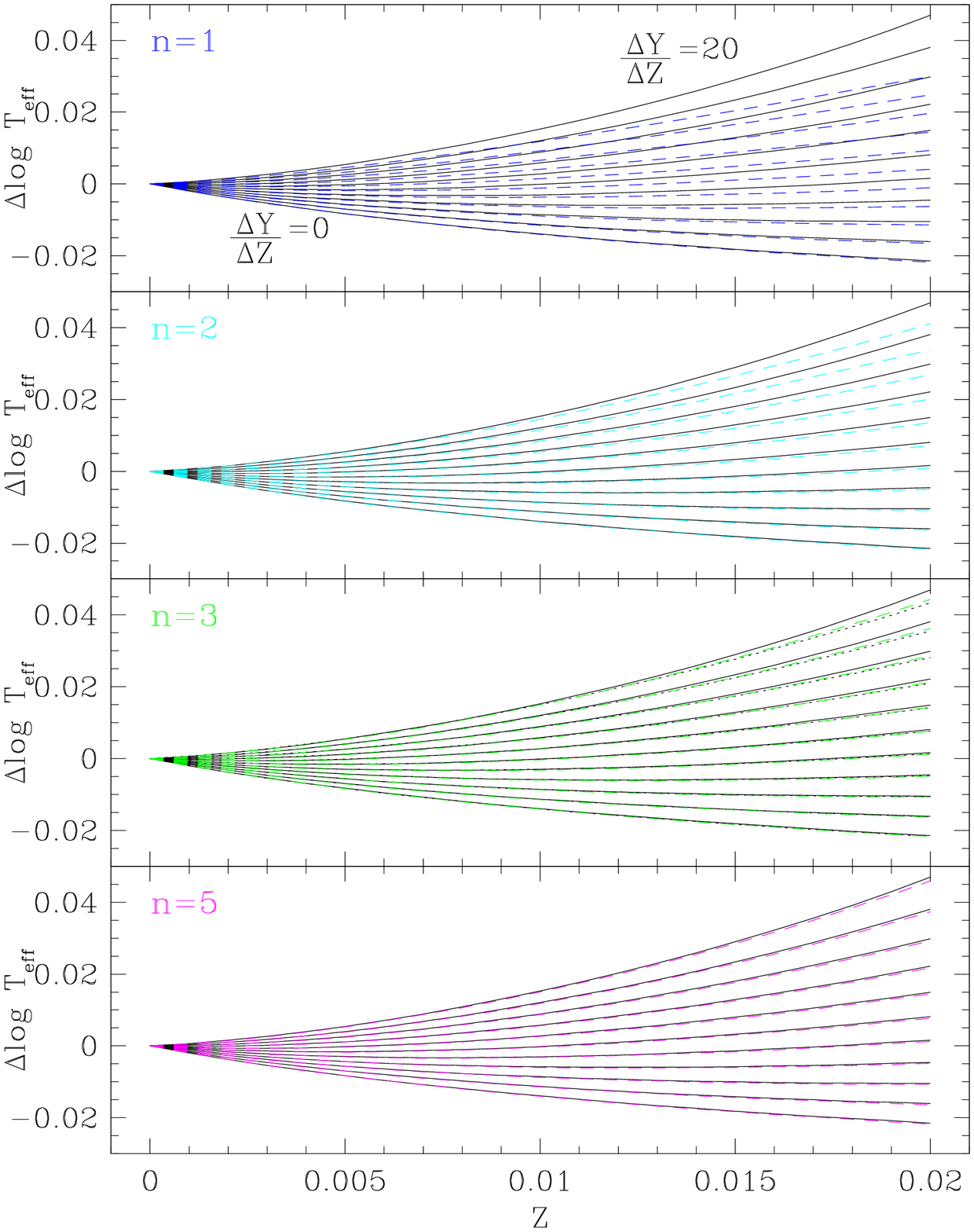}
\includegraphics[scale=0.43]{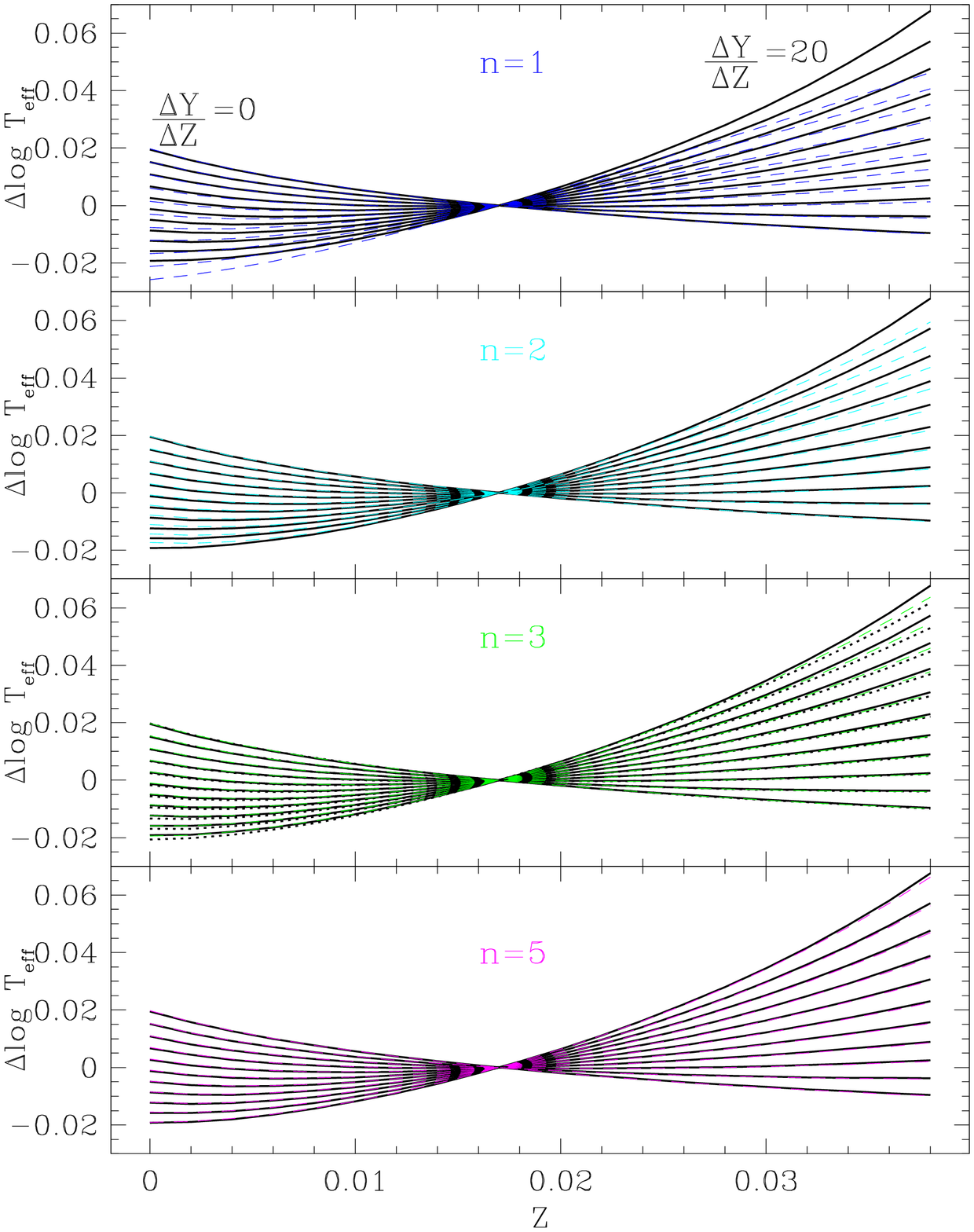}
\caption{Homology relations for $\frac{\Delta Y}{\Delta Z}$ ranging 
from 0 to 20 (with a pace of 2) as indicated. Coloured dashed lines
represent the approximate relations of Eq.~\protect{\ref{eq:polyn_approx}} 
for increasing order of the terms of the expansion.
The true relations are well reproduced up to $\frac{\Delta Y}{\Delta Z}=20$
for $n>2$ --- at least within a metallicity range 
$\Delta Z \stackrel{\textstyle <}{\sim} 0.02$. The dotted line shows
our simplified form of the homology relations 
(Eq.~\ref{eq:DlogTeff_2terms}).}
\end{center}
\label{fig:homology_approx}
\end{figure*}
%%%%%%%%%%%%%%%%%%%%%%%

%%%%%%%%%%%%%%%%%%%%%%%%%%%%%%%%%%%%%%%%%%%%%%%%%%%%%%%%%%%%%%%%%%%%%%%%%%%%%%
\subsection[]{Polynomial approximations}

Pagel \& Portinari (1998) suggested an approximation of the homology relations 
based on Taylor series expansion of the terms in $\log (1-x)$:
\[ \log (1-x) = 
    -\frac{1}{\ln 10} \sum_{n=1}^\infty \frac{x^n}{n} ~~~~~~~~~~~~~~~~~~|x| <1 \] 
Expansion to first order of Eq.~\ref{eq:DlogTeff} yields a linearized
formula similar to eq.~5 of Pagel \& Portinari:
\begin{equation} 
\label{eq:linear_approx}
\begin{array}{l}
\medskip
\Delta \log {\teff}^{(1)} = 
                        -0.051 \log \left( \frac{100 Z+1}{100 Z_r+1} \right) \\

\medskip
\;\;\;\;\;\;\;\;\;\;\;\;\;\; + \left[ \frac{0.0406}{X_r}+ \frac{0.02215}{1+X_r}+
                      \frac{0.4256}{3+5 X_r- Z_r} \right] \delta (Z-Z_r) \\
                       
\;\;\;\;\;\;\;\;\;\;\;\;\;\; +  \frac{0.08512}{3+5 X_r- Z_r} (Z-Z_r) \\
\end{array} 
\end{equation}
(where the last term is negligible). However this linearized approximation 
holds only for $| \delta (Z-Z_r) | << 1$ and does not recover very well the 
original homology relations for  $\frac{\Delta Y}{\Delta Z} > 5$ 
(see Fig.~A1). 

One can then resort to further terms in the expansion. The second order
terms yield:
\[ \begin{array}{l}
\medskip
\Delta \log {\teff}^{(2)} = \Delta \log {\teff}^{(1)}\\

\medskip
\;\;\;\;\;\;\;\;\;\;\;\;\;\;\;  
                       + \left[ \frac{0.0203}{X_r^2} + \frac{0.01107}{(1+X_r)^2}
+ \frac{1.064}{(3+5 X_r- Z_r)^2} \right] \delta^2 (Z-Z_r)^2 \\

\medskip
\;\;\;\;\;\;\;\;\;\;\;\;\;\;\;  
                    + \frac{0.4256}{(3+5 X_r- Z_r)^2} \, \delta \, (Z-Z_r)^2 \\

\;\;\;\;\;\;\;\;\;\;\;\;\;\;\;
                              + \frac{0.04256}{(3+5 X_r- Z_r)^2} \, (Z-Z_r)^2 \\
\end{array} \]
where the last two terms are negligible. The third order expansion terms yield:
\[ \begin{array}{l}
\medskip
\Delta \log {\teff}^{(3)} = \Delta \log {\teff}^{(2)} + \\

\medskip
\;\;\;\;\;\;\;\;\;\;\;\;\;\;\;  
                  \left[ \frac{0.01354}{X_r^3} + \frac{0.00738}{(1+X_r)^3} +
              \frac{3.547}{(3+5 X_r- Z_r)^3} \right] \delta^3 (Z-Z_r)^3 \\

\medskip
\;\;\;\;\;\;\;\;\;\;\;\;\;\;\;  
                  + \frac{0.2128}{(3+5 X_r- Z_r)^3} \, \delta^2 \, (Z-Z_r)^3 \\

\medskip
\;\;\;\;\;\;\;\;\;\;\;\;\;\;\;  
                    + \frac{0.4256}{(3+5 X_r- Z_r)^3} \, \delta \, (Z-Z_r)^2 \\

\;\;\;\;\;\;\;\;\;\;\;\;\;\;\;  
                        + \frac{0.0284}{(3+5 X_r- Z_r)^3} \, (Z-Z_r)^3 \\
\end{array} \]
where the last three terms are negligible. In general, neglecting
the smaller terms of the kind $\delta^m (Z-Z_r)^n$ with $m<n$, we can write
the following polynomial approximation to homology relations:
\begin{equation} 
\label{eq:polyn_approx}
\begin{array}{l}
\medskip
\Delta \log \teff \simeq -0.051 \log \left( \frac{100 Z+1}{100 Z_r+1} 
                             \right) \\

                          + \sum_{n=1}^\infty \left[ \frac{0.0406}{X_r^n} + 
                            \frac{0.02215}{(1+X_r)^n} +
                            \frac{0.08512}{(3+5 X_r- Z_r)^n} \right] 
                            \frac{\delta^n (Z-Z_r)^n}{n} \\
\end{array} 
\end{equation}
The corresponding parametric fitting formula for the isochrones would be
of the kind:
\begin{equation}                
\label{eq:EQ4}
\Delta \log \teff \simeq -a_1 \log \left( \frac{a_2 Z+1}{a_2 Z_r+1} \right)+
                            \sum_{n=1}^\infty b_n \delta^n (Z-Z_r)^n
\end{equation}                
For $n$-th order series expansion, this fitting formula has $n+2$ free 
parameters. Pagel \& Portinari (1998) indeed used a linearized approximation 
($n=1$) similar to Eq.~\ref{eq:linear_approx} as a guideline, and had 
3 parameters in their fitting formula for isochrones.

Fig.~A1 shows that expansion to order $n>2$ 
is needed to yield a good fit up to $\Delta Y/\Delta Z \sim 10$ (at least 
in the range $0 \leq Z$ \lsim $Z_\odot$) needed for the study of nearby 
K dwarfs of low Z. Our simplified form of the original homology relations 
(Eq.~\ref{eq:DlogTeff_2terms}) performs as well as the $n=3$ order of series
expansion, with just 3 free parameters in the corresponding fitting formula.

For extremely high $\Delta Y/\Delta Z$, as applies to the case of $\omega$Cen,
it is worth using the original formulation of the homology relations 
(Eq.~\ref{eq:DlogTeff}), which indeed we showed in Section~\ref{sec:homology_GC}
to yield excellent results when applied to the multiple MS of $\omega$Cen
amd NGC 2808.
Notice, in fact, that for very high $\Delta Y/\Delta Z$ the basic
condition for Taylor series expansion, {\mbox{$|x| = |\delta (Z-Z_r)| < 1$}}, 
may no longer be fulfilled.

%%%%%%%%%%%%%%%%%%%%%%%%%%%%%%%%%%%%%%%%%%%%%%%%%%%%%%%%%%%%%%%%%%%%%%%%%%%%%%

\section[]{Updated Hipparcos parallaxes}
For simplicity this paper relies on the helium abundances determined
by Casagrande et al.\ (2007); in the meantime, the HR diagram of nearby stars 
has been revised with updated Hipparcos parallaxes (van Leewven 2007).
In Fig.~B1 we show that the effect of the revised parallaxes
is minimal: the sample only slightly changes (3 stars are now rejected 
by the parallax limit 6\%, while 5 new stars are included, for a total of 88 
objects) and the trend toward significantly sub-cosmological Y values is
confirmed. Nor it appears to be due to low metallicity stars being
systematically more distant than the solar metallicity ones
(with systematically intrinsic bluer colours due to neglected reddening):
though a slight distance--metallicity trend is present as expected,
most of the stars lie within 30 pc for all metallicities 
(Fig.~B2). Also, in Fig.~\ref{fig:WCen} we show that the reddening/extinction
correction needed to reconcile metal-poor stars with isochrones would 
correspond to $E(B-V)$=0.10, which is certainly too extreme for so nearby stars.

Other possible systematics have been carefully 
checked and excluded in Casagrande et al.\ (2007). We do not deem necessary 
here to repeat the detailed Monte--Carlo simulations performed in 2007 to  
estimate realistic errorbars (shown in Fig.~\ref{fig:YZfield}), 
since it suffices here to demonstrate that the new parallaxes by no means 
solve the riddle of sub-cosmological helium abundances.

Finally, we have also verified that the updated temperature scale 
by Casagrande et al.\ (2010) also does not significantly change the Y(Z) plot.

%%%%%%%%%%%%%%%%%%%%%%% Figure B1
\begin{figure}
\begin{center}
\includegraphics[scale=0.32,angle=-90]{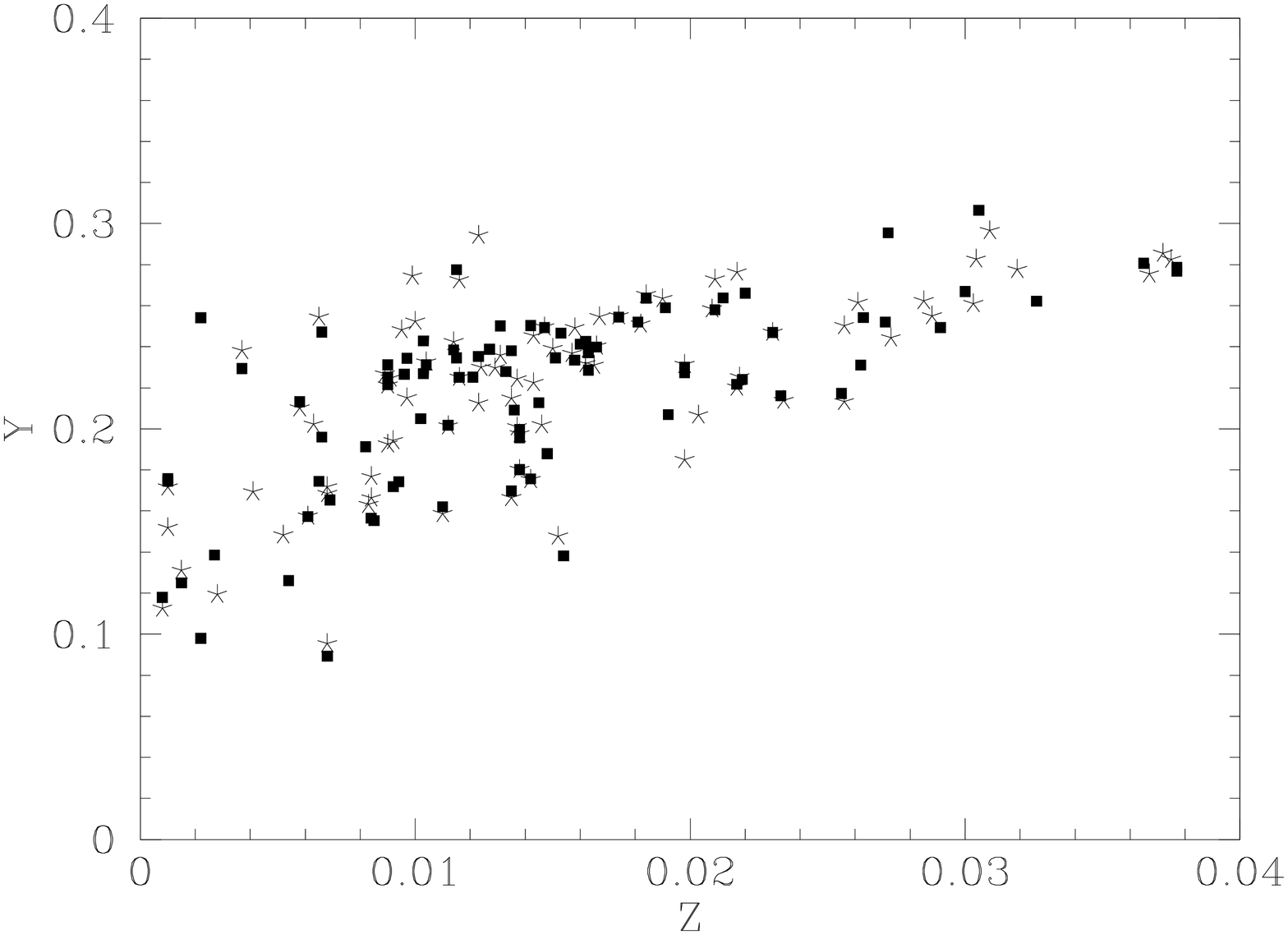}
\caption{Helium plot for the sample by Casagrande et al.\ (2007) 
({\it stars}) versus the update sample with the revised parallaxes of
van Leewven (2007) ({\it squares})}
\end{center}
\label{fig:vanLeewven}
\end{figure}
%%%%%%%%%%%%%%%%%%%%%%%

%%%%%%%%%%%%%%%%%%%%%%% Figure B2
\begin{figure}
\begin{center}
\includegraphics[scale=0.4]{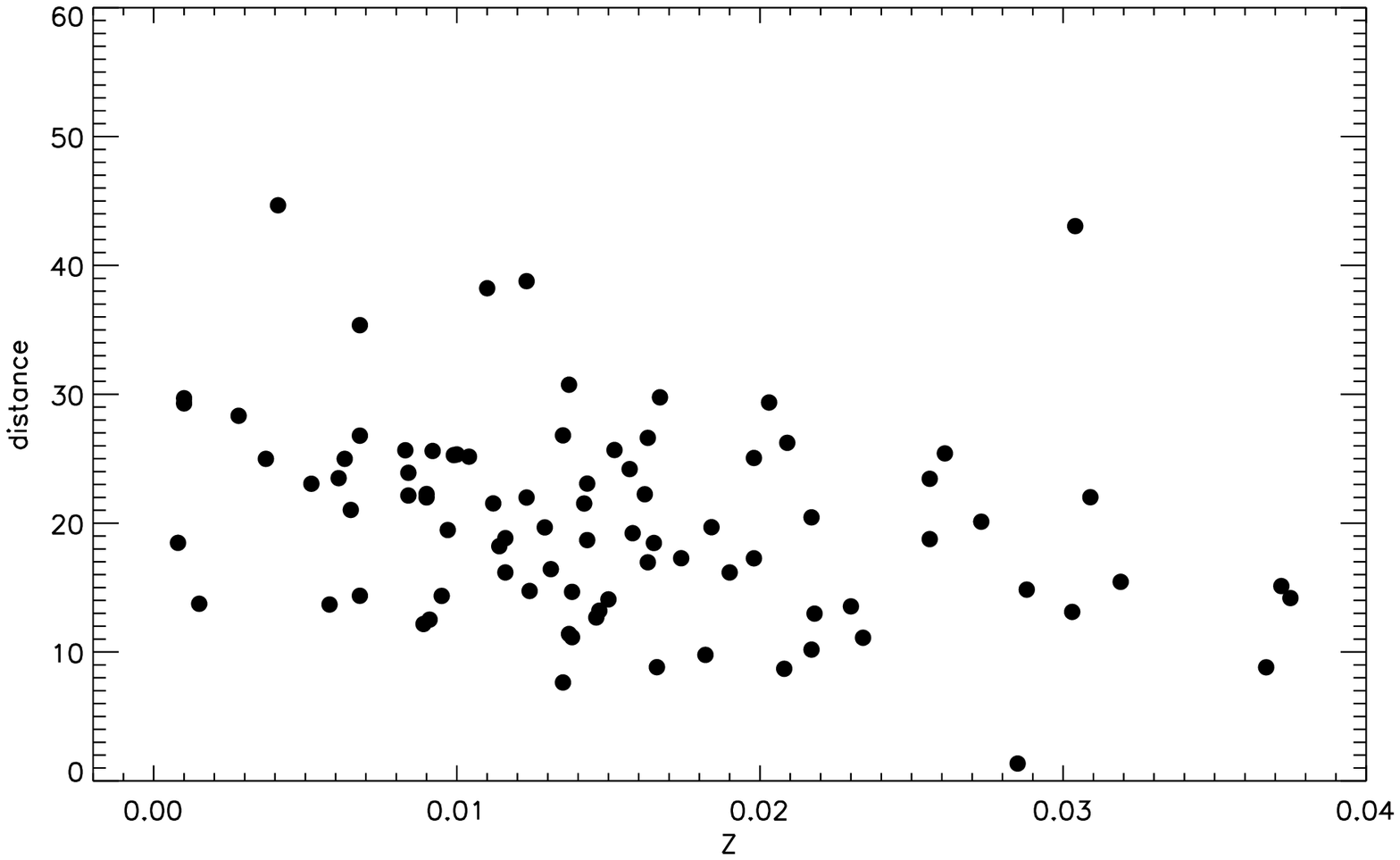}
\includegraphics[scale=0.4]{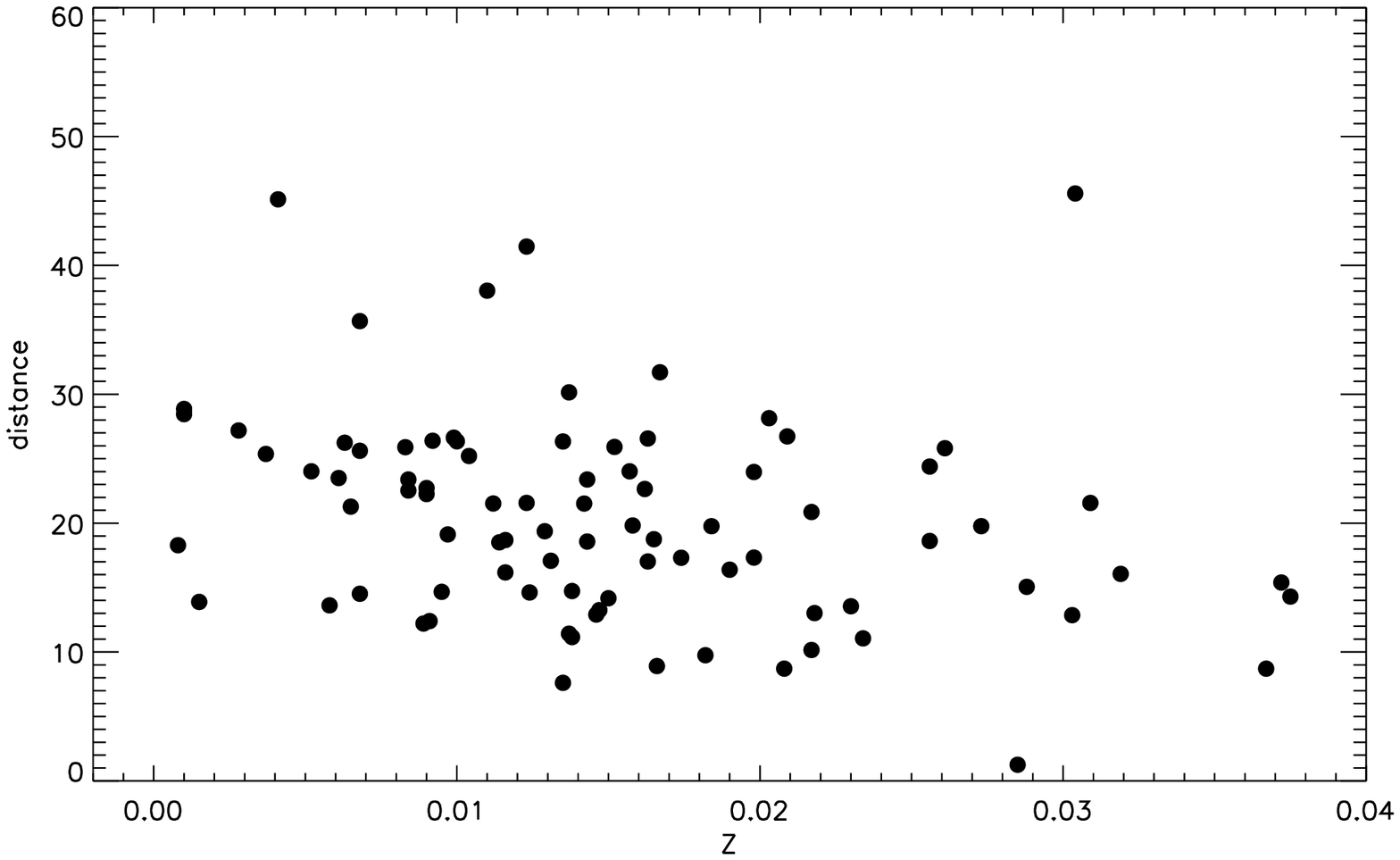}
\caption{Distances and metallicities of sample stars from Casagrande
et~al.\ (2007) ({\it top panel}) and for the updated sample with the
revised parallaxes of van Leewven (2007)({\it bottom panel}) } 
\end{center}
\label{fig:distances}
\end{figure}
%%%%%%%%%%%%%%%%%%%%%%%


\begin{thebibliography}{}
\bibitem{}
Alonso A., Arribas S., Martinez-Roger C., 1996, A\&AS 117, 227
\bibitem{}
Anderson J., Piotto G., King I.R., Bedin L.R., Guhathakurta P., 2009, 
ApJ 697, L58
\bibitem{}
Asplund M., Grevesse N., Sauval A.J., Scott P., 2009, ARA\&A, 47, 481
\bibitem{}
Balser D.S., 2006, AJ 132, 2326
\bibitem{}
Basu S., Antia H.M., 2008, Phys.\ Rep.\ 457, 217
\bibitem{}
Bedin L.R., Piotto G., Anderson J., Cassisi S., King Y.R., Momany Y., 
Carraro G., 2004, ApJ 605, 125
\bibitem{}
Bellazzini M., Ferraro F.R., Sollima A., Pancino E., Origlia L., 2004, 
A\&A 424, 199 
\bibitem{}
Bertelli G., Girardi L., Marigo P., Nasi E., 2008, A\&A 484, 815
\bibitem{}
Boyajian T.S., McAlister H.A., Baines E.K., \etal 2008, ApJ 683, 424
\bibitem{}
Caloi V., D'Antona F., 2005, A\&A 435, 987
\bibitem{}
Caloi V., D'Antona F., 2007, A\&A 463, 949
\bibitem{}
Carigi L., Peimbert M., 2008, RMxAA 44, 341
\bibitem{}
Carretta E., Bragaglia A., Gratton R.G., Leone F., Recio-Blanco A., 
Lucatello S., 2006, A\&A 450, 523
\bibitem{}
Carretta E., Bragaglia A., Gratton R.G., D'Orazi V., Lucatello S., 2009, 
A\&A in press (arXiv:0910.0675)
\bibitem{}
Casagrande L., 2008, PhD Thesis, University of Turku
\bibitem{}
Casagrande L., Portinari L., Flynn C., 2006, MNRAS 373, 13
\bibitem{}
Casagrande L., Flynn C., Portinari L., Girardi L., Jimenez J., 2007, 
MNRAS 382, 1516
\bibitem{}
Casagrande L., Ram\'irez I., Mel\'endez J., Bessel M., Asplund M., 2010, 
A\&A in press (arXiv:1001.3142)
\bibitem{}
Castellani V., Degl'Innocenti S., Marconi M., 1999, A\&A 349, 834
\bibitem{}
Catelan M., Valcarce A.A.R., Sweigart A.V., 2009a, IAU Symp. 266, eds.\ 
R.\ de Grijs and J.R.D.\ Lepine (arXiv:0910.1367)
\bibitem{}
Catelan M., Grundahl F., Sweigart A.V., Valcarce A.A.R., Cort\'es C., 2009b, 
ApJ 695, L97
\bibitem{}
Chaboyer B., Fenton W.H., Nelan J.E., Patnaude D.J., Simon F.E., 2001,
ApJ 562, 521
\bibitem{}
Chiosi C., Matteucci M., 1982, A\&A 105, 140
\bibitem{}
Choi E., Yi S.-K., 2007, MNRAS 375, L1
\bibitem{}
Choi E., Yi S.-K., 2008, MNRAS 386, 1332 
\bibitem{}
Clausen J.V., Olsen E.H., Helt B.E., Claret A., 2010, A\&A in press 
(arXiv:0912.3108)
\bibitem{}
Cox J.P., Giuli R.T., 1968, Principles of Stellar Structure - Volume II. 
Gordon \& Breach Science Publishers, New York
\bibitem{}
Cox J.P., Giuli R.T., 2004, Principles of Stellar Structure. Extended Second
Edition by A.\ Weiss, W.\ Hillebrandt, H.-C.\ Thomas and H.\ Ritter, Cambridge
Scientific Publishers, Cambridge, UK
\bibitem{}
D'Antona F., Caloi V., 2004, ApJ 611, 871
\bibitem{}
D'Antona F., Bellazzini M., Caloi V., Pecci F.F., Galleti S., Rood R.T., 2005, 
ApJ 631, 868
\bibitem{}
Del Principe M., Piersimoni A.M., Storm J., \etal 2006, ApJ 652, 362
\bibitem{}
Faulkner J., 1967, ApJ 147, 617
\bibitem{}
Fernandes J., Lebreton Y., Baglin A., 1996, A\&A 311, 127
\bibitem{}
Fernandes J., Lebreton Y., Baglin A., Morel P., 1998, A\&A 338, 455
\bibitem{}
Fukugita M., Kawasaki M., 2006, ApJ 646, 691
%\bibitem{}
%Gonz\'alez Hern\'andez J.I., Bonifacio P., 2009, A\&A 497, 497
\bibitem{}
Harris W.E., 1996, AJ 112, 1487
\bibitem{}
Iglesias J.A., Rogers F.J., 1996, ApJ 464, 943
\bibitem{}
Izotov Y.I., Thuan T.X., Stasinska G., 2007, ApJ 662, 15
\bibitem{}
Jimenez R., Flynn C., McDonald J., Gibson B., 2003, Science 299, 1552
\bibitem{}
Karakas A., Fenner Y., Sills A., Campbell S.W., Lattanzio J., 2006, 
ApJ 652, 1240
\bibitem{}
Korn A.J., Grundahl F., Richard O., Barklem P.S., Mashonkina L., Collet, R., 
Piskunov N., Gustafsson B., 2006, Nat 442, 657
\bibitem{}
Korn A. J., Grundahl F., Richard O., Mashonkina L., Barklem P.S., Collet R., 
Gustafsson B., Piskunov N., 2007, ApJ 671, 402
\bibitem{}
Lebreton Y., Perrin M.-N.,, Cayrel R., Baglin A., Fernandes J., 1999, 
A\&A 350, 587
\bibitem{}
Lee Y.-W., Joo S.-J., Han S.-I., \etal 2005, ApJ 621, L57
\bibitem{}
Lee J.-W., Lee J., Kang Y.-W., \etal 2009, ApJ 695, L78
\bibitem{}
van Leeuwen F., 2007, A\&A 474, 653
\bibitem{}
Lub J., 2002, in $\omega$ Centauri: a unique window into astrophysics, ed.
F.\ van Leeuwen, J.\ Hughes, G.\ Piotto, (San Francisco:ASP), ASP Conf.\ Ser.\
265, p.\ 95
\bibitem{}
Marcolini A., Gibson B.K., Karakas A.I., S\'anchez--Bl\'azquez P., 2009, 
MNRAS 395, 719
\bibitem{}
Maeder A., 1992, A\&A 264, 105
\bibitem{}
Maeder A., Meynet G., 2006, A\&A 448, L37
\bibitem{}
Michaud G., Vauclair G., Vauclair S., 1983, ApJ 267, 256
\bibitem{}
Michaud G., Richer J., Richard O., 2008, ApJ 675, 1223
\bibitem{}
Milone A.P., Piotto G., King I.R., et~al.\ 2010, ApJ 709, 1183
\bibitem{}
Norris J.E., 2004, ApJ 612, 25
\bibitem{}
Nordstr\"om B., Mayor M., Andersen J., et~al.\ 2004, A\&A 418, 989
\bibitem{}
Pagel B.E.J., Portinari L., 1998, MNRAS 298, 747
\bibitem{}
Peimbert M., 2003, ApJ 584, 735
\bibitem{}
Peimbert M., Luridiana V., Peimbert A., 2007, ApJ 666, 636
\bibitem{}
Peng F., Nagai D., 2009, ApJL 705, L58
\bibitem{}
Perrin M.-N., de Strobel G.C., Cayrel R., Hejlesen P.M., 1977, A\&A 54, 779 
\bibitem{}
Piotto G., 2009, IAU Symp.\ 258 in press (arXiv:0902.1422)
\bibitem{}
Piotto G., Villanova S., Bedin L.R., Gratton R., Cassisi S., et al. 2005,
ApJ 621, 777
\bibitem{}
Piotto G., Bedin L.R., Anderson J., King I.R., Cassisi S., et al. 2007,
ApJ 661, L53
\bibitem{}
Prantzos N., Charbonnel C., 2006, A\&A 458, 135
\bibitem{}
Ram\'irez I., Mel\'endez J., 2005, ApJ 626, 446
\bibitem{}
Renzini A., 2008, MNRAS 391, 354
\bibitem{}
Richard O., Michaud G., Richer J., 2002, ApJ 580, 1100
\bibitem{}
Richard O., Michaud G., Richer J., 2005, ApJ 619, 538
\bibitem{}
Romano D., Matteucci F., Tosi M., Pancino E., Bellazzini M., Ferraro F.,
Limongi M., Sollima A., 2007, MNRAS 376, 405
\bibitem{}
Romano D., Tosi M., Cignoni M., Matteucci F., Pancino E., Bellazzini M., 2009, 
MNRAS in press (arXiv:0910.1299)
\bibitem{}
Salaris M., Weiss A., Ferguson J.W., Fusilier D.J., 2006, ApJ 645, 1131
\bibitem{}
Seaton M.J., 2005, MNRAS 362, L1
\bibitem{}
Serenelli A., Basu S., Ferguson J.W., Asplund M., 2009, ApJ 705, L123
\bibitem{}
Sirianni M., Jee M.J., Benítez N., \etal 2005, PASP 117, 1049
\bibitem{}
Sollima A., Ferraro F.R., Pancino E.,, Bellazzini M., 2005, MNRAS 357, 265
\bibitem{}
Sollima A., Borissova J., Catelan M., Smith H.A., Minniti D., Cacciari C., 
Ferraro F.R., 2006, ApJ 640, L43
\bibitem{}
Sollima A., Ferraro F.R., Bellazzini M., Origlia L., Straniero O., Pancino E.,
2007, ApJ 654, 915
\bibitem{}
Sousa S.G., Santos N.C., Mayor M., 2008, A\&A 487, 373
\bibitem{}
Steigman G., 2007, Annual Review of Nuclear and Particle Systems, 57, 463
\bibitem{}
Torres G., Boden A., Latham D.W., Pan M., Stefanik R.P., 2002, AJ, 124, 1716
\bibitem{}
VandenBerg D.A., Edvardsson B., Eriksson K., Gustafsson B., 2008, 
ApJ 675, 746
\bibitem{}
Villanova S., Piotto G., Gratton R.G., 2009, A\&A in press (arXiv:0903.3924) 
\bibitem{}
Villanova S., Piotto G., King I.R., Anderson J., Bedin L.R., et al. 2007,
ApJ 663, 296
\bibitem{}
Yi S.-K., 2009, IAU Symp.\ 258, eds.\ 253
\bibitem{}
Yi S., Demarque P., Kim Y.C., Lee Y.W., Ree C.H., Lejeune T., Barnes S., 2001,
ApJS, 136, 417
\end{thebibliography}
\end{document}